\documentclass[12pt,habil,bindborder,english,secnumarabic,eqsecnum,nofootinbib]{tuthesis}
\usepackage{graphicx}
\usepackage{amssymb,amsfonts,amsmath,bm,theorem}
\usepackage[dvips,colorlinks=true,linkcolor=blue]{hyperref}
\usepackage{mathrsfs}
\usepackage{wrapfig}
\newcommand{\pard}[2]{\frac{\partial #1}{\partial #2}}

\newcommand{\bc}{\begin{center}}
\newcommand{\ec}{\end{center}}
\newcommand{\be}{\begin{equation}}
\newcommand{\ee}{\end{equation}}
\newcommand{\ba}{\begin{eqnarray*}}
\newcommand{\ea}{\end{eqnarray*}}
\newcommand{\bna}{\begin{eqnarray}}
\newcommand{\ena}{\end{eqnarray}}

\newcommand{\mpaa}{\begin{minipage}[t]{7.5cm}}
\newcommand{\mpea}{\end{minipage}}

{\theorembodyfont{\rmfamily}\theoremstyle{plain}

}
{\theorembodyfont{\rmfamily}\theoremstyle{break}
}
\newcommand{\qed}{\phantom{xxxxxx}\hfill q.e.d.}

\begin{document}

\bc
{\Large \bf Weak chaos, infinite ergodic theory,\\[1ex]
and anomalous dynamics}\\[2ex]
{Rainer Klages\\
Queen Mary University of London\\
School of Mathematical Sciences\\
Mile End Road, London E1 4NS\\[1ex]
e-mail: r.klages@qmul.ac.uk}
\ec

\begin{abstract}
This book chapter introduces to the concept of weak chaos, aspects of
its ergodic theory description, and properties of the anomalous
dynamics associated with it. In the first half of the chapter we study
simple one-dimensional deterministic maps, in the second half basic
stochastic models and eventually an experiment. We start by reminding
the reader of fundamental chaos quantities and their relation to each
other, exemplified by the paradigmatic Bernoulli shift. Using the
intermittent Pomeau-Manneville map the problem of weak chaos and
infinite ergodic theory is outlined, defining a very recent
mathematical field of research. Considering a spatially extended
version of the Pomeau-Manneville map leads us to the phenomenon of
anomalous diffusion. This problem will be discussed by applying
stochastic continuous time random walk theory and by deriving a
fractional diffusion equation. Another important topic within modern
nonequilibrium statistical physics are fluctuation relations, which we
investigate for anomalous dynamics. The chapter concludes by showing
the importance of anomalous dynamics for understanding experimental
results on biological cell migration.
\end{abstract}

  
\newpage
\setcounter{tocdepth}{1}
\tableofcontents

\newpage
  

\chapter{Introduction}

Deterministic dynamical systems involving only a few variables can
exhibit {\em complexity} reminiscent of many-particle systems if the
dynamics is {\em chaotic}, as is quantified by the existence of a
positive Lyapunov exponent \cite{Schu,Ott,Beck,ASY97}. Such systems,
which may be called {\em small} because of their small number of
degrees of freedom \cite{KJJ12}, can display an intricate interplay
between nonlinear microscopic dynamical properties and macroscopic
statistical behavior leading to highly non-trivial fluctuations of
physical observables. This becomes particularly interesting in
nonequilibrium situations when these systems are exposed to external
gradients or fields. Despite their complexity, examples of these
systems are still amenable to detailed analysis by means of dynamical
systems theory in combination with stochastic theory. Hence, they
provide important paradigms to construct a theory of nonequilibrium
statistical physics from first principles: Based on the chaotic
hypothesis, which generalizes Boltzmann's ergodic hypothesis, SRB
measures were studied as nonequilibrium equivalents of the Gibbs
ensembles of equilibrium statistical mechanics. This novel approach
led to the discovery of fundamental relations characterizing
nonequilibrium transport in terms of microscopic chaos
\cite{Do99,Gasp,Kla06,CFLV08}, such as formulas expressing transport
coefficients in terms of Lyapunov exponents and dynamical entropies,
equations relating the nonequilibrium entropy production to the
fractality of SRB measures, and fluctuation relations, which are now
widely studied as a fundamental property of nonequilibrium processes
\cite{EvSe02,BLR05,Kla06,KJJ12}.

\begin{figure}
\centerline{\includegraphics[width=0.38\textwidth,angle=-90]{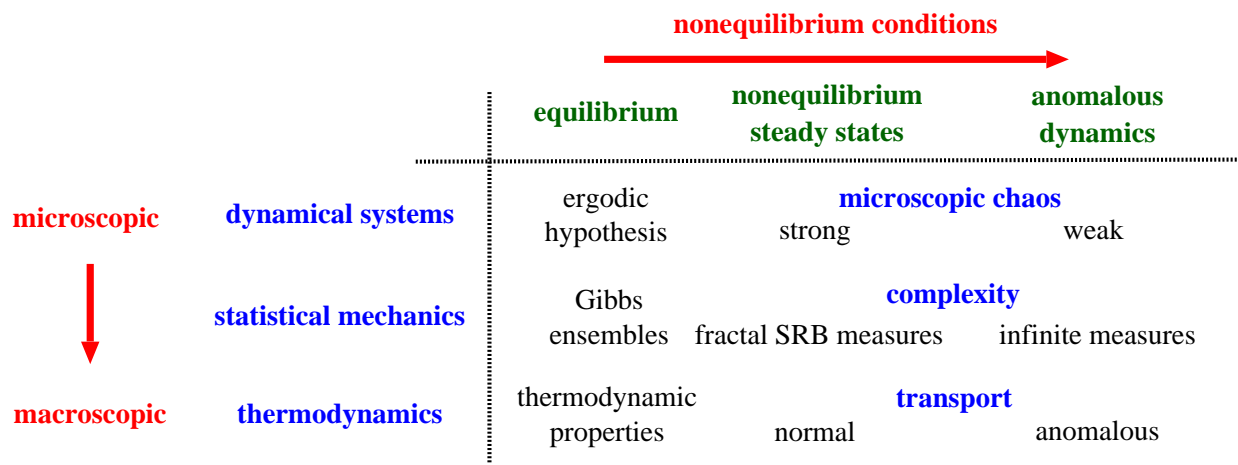}}
\caption{Conceptual foundations of a theory of nonequilibrium statistical 
physics based on dynamical systems theory by motivating the topic of
this book chapter, which is represented by the third column.}
\label{fig:bigp} 
\end{figure}

The interplay between these different levels of description in modern
nonequilibrium statistical mechanics is illustrated by the second
column in Fig.~\ref{fig:bigp}, in analogy to the theory of equilibrium
statistical mechanics sketched in the first column. As is represented
by the third column, however, more recently scientists learned that
random-looking evolution in time and space also occurs under
conditions that are weaker than requiring a positive Lyapunov exponent
\cite{Aar97,ZaUs01}. It is now known that there is a wealth of systems
exhibiting zero Lyapunov exponents, meaning that the separation of
nearby trajectories is weaker than exponential. This class of
dynamical systems is called {\em weakly chaotic}. Examples include
maps with indifferent fixed points, polygonal particle billiards, and
Hamiltonian systems with sticky islands in phase space
\cite{Aar97,ZaUs01,Kla06,KRS08}. 

Weakly chaotic systems exhibit {\em anomalous dynamics} characterized
by novel properties such as ageing, which reflects an extremely weak
relaxation towards equilibrium involving more than one time scale in
the decay of correlations. Other surprising properties are the
existence of L\'evy-type probability distributions obeying generalized
central limit theorems \cite{SZK93,KSZ96} and the non-equivalence of
time and ensemble averages, called weak ergodicity breaking
\cite{SHB09}. These physical phenomena were observed
experimentally in a wide variety of systems, such as in the anomalous
statistics of blinking quantum dots, in the anomalous diffusion of
atoms in optical lattices, in plasma physics and even in cell and
animal migration \cite{MeKl00,MeKl04,KRS08,SHB09}.

Recent work in ergodic theory, on the other hand, has led to
mathematically rigorous results about some of the physically relevant
phenomena mentioned above. It turned out that there is an intimate
connection between the mechanism generating weakly chaotic dynamics
and the existence of non-normalizable, so-called {\em infinite
invariant measures} \cite{Aar97,Aar00,ThZw06}. The ergodic theory of
generalized random walks driven by weak chaos and of other systems
exhibiting infinite invariant measures, which is called {\em infinite
ergodic theory}, has thus the potential of providing a sound
mathematical basis for some of the physical phenomena displayed by
anomalous dynamics.

This book chapter gives a brief introduction to important aspects of
the above topics in four sections: As a warm-up, the beginning of
Section~\ref{sec:cad} briefly reminds us of the concept of
deterministic chaos in simple dynamical systems as quantified by a
positive Lyapunov exponent. On this basis, we will introduce to the
phenomenon of weak chaos, and the idea of infinite ergodic theory will
be outlined. The chapter concludes by putting different forms of
deterministic chaos into perspective. Section~\ref{sec:anodif} relates
these concepts and ideas to the problem of anomalous diffusion in
deterministic systems. Here we make a transition to stochastic theory
by studying these systems also from a stochastic point of view. For
this purpose we use a generalization of ordinary random walk theory,
called continuous time random walk theory. In a scaling limit, this
theory leads to generalized diffusion equations involving fractional
derivatives. Section~\ref{sec:afrc} introduces to the topic of
fluctuation relations, which generalize the Second Law of
Thermodynamics and other fundamental thermodynamic relations to small
systems far away from equilibrium. After discussing transient
fluctuation relations for a very basic type of stochastic dynamics as
an example, we explore the validity of such relations for
generalizations of this dynamics yielding anomalous diffusion. In
section~\ref{sec:cell} we relate this line of theoretical reasoning
about anomalous dynamics to biophysical reality by studying the case
of biological cell migration. After briefly introducing to the problem
of cell migration, we report experimental results on fundamental
statistical physical properties of migrating cells, extracted from
statistical data analysis. We conclude this section with a stochastic
modeling of these experimental results by using a generalized,
fractional Fokker-Planck type equation. We summarize our discussion of
this book chapter in the final section~\ref{sec:summ}.

The title of this review is inspired by a conference that the author
had the pleasure to organize together with R.Zweim\"uller, E.Barkai,
and H.Kantz at the Max Planck Institute for the Physics of Complex
Systems, Dresden, in Summer 2011, which bears exactly the same title
\cite{wchaos11}. However, naturally this chapter reflects the author's
very personal take on this topic, and his own research. The subsequent
second section is to some extent based on the review Ref.~\cite{Kla10}
by combining it with ideas from Refs.~\cite{Kla06,HoKl}. The third
section builds on Refs.~\cite{KCKSG06,KKCSG07}. The fourth section
incorporates material from the review Ref.~\cite{KCD12} and from
Ref.~\cite{ChKl09}, the fifth one draws on Ref.~\cite{DKPS08}.

\chapter{Chaos and anomalous dynamics}\label{sec:cad}

In this section we focus on purely deterministic dynamics modeled by
two simple but paradigmatic one-dimensional maps: the famous Bernoulli
shift, as a model for strong chaos characterized by a positive
Lyapunov exponent, and the Pomeau-Manneville map, as an example
exhibiting weak chaos with zero Lyapunov exponent. We start by briefly
reminding of basic concepts of dynamical systems theory and ergodic
theory such as Lyapunov exponents, ergodicity, SRB measures, and
Pesin's theorem, illustrated for the Bernoulli shift.
Ref.~\cite{Kla10} provides a more tutorial exposition of most of these
ideas. By switching to the Pomeau-Manneville map we find that
generalizations of these concepts are needed in order to describe the
model's weakly chaotic dynamics. This motivates the mathematical
problem of infinite ergodic theory, which is intimately related to
defining suitably generalized chaos quantities assessing weak chaos,
and a generalization of Pesin's theorem. In the final part of this
chapter we propose a generalized hierarchy of chaos, based on the
existence of different types of stretching between two nearby
trajectories, which we use to characterize chaotic dynamics.

\section{Deterministic chaos in a simple map}\label{sec:chsima}

The main vehicle of our approach in this and the next section are
one-dimensional time-discrete maps $F:J\to J\:,\:J\subseteq
\mathbb{R}$ obeying
\be
x_{n+1}=F(x_n)\:,\: n\in\mathbb{N}_0 \: , \label{eq:eom}
\ee 
which defines the equations of motion of our deterministic dynamical
systems. For a given initial condition $x_0$ we have $x_n=F^n(x_0)$. A
particularly simple example of $F$ are piecewise linear maps, such as
the paradigmatic Bernoulli shift \cite{Schu,Ott,ASY97,Do99}
\begin{figure}[t]
\centerline{\includegraphics[height=7cm]{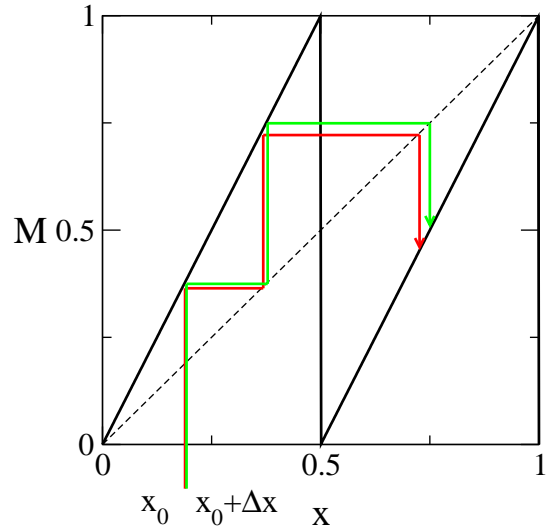}}
\caption{The Bernoulli shift Eq.~(\ref{eq:bern}) and two trajectories 
starting from two nearby initial conditions $x_0$ and $x_0'=x_0+\Delta
x_0$ displaced by $\Delta x_0\ll 1$.}
\label{fig:bern} 
\end{figure}
\be
B:[0,1)\to[0,1)\;,\;B(x):=2x \;\mbox{mod 1}\; = \left\{\begin{array}{l} 2x\;,\; 0\le x < 1/2 \\ 
2x-1\;,\; 1/2 \le x < 1 \end{array}\right. \label{eq:bern}
\ee
depicted in Fig.~\ref{fig:bern}.  This simple system exhibits a very
complicated dynamics governed by sensitivity to initial conditions, as
can be quantified by calculating its Lyapunov exponent
\cite{Ott,Rob95}: Consider two points that are initially displaced
from each other by $\Delta x_0 :=|x_0'-x_0|$ with $\Delta x_0$
``infinitesimally small'' such that $x_0,x_0'$ do not hit different
branches of the Bernoulli shift $B(x)$ around $x=1/2$.\footnote{This
condition could be eliminated by defining a metric on a circle
\cite{ASY97}.} We then have
\be
\Delta x_n := |x_n'-x_n|=2\Delta x_{n-1} = 2^2\Delta x_{n-2}
=\ldots=2^n\Delta x_0 =e^{n \ln2} \Delta x_0 \quad . \label{eq:lyapbern}
\ee
We thus see that there is an exponential separation between two nearby
points as we follow their trajectories, where the rate of separation
$\lambda(x_0):=\ln 2$ is the (local) Lyapunov exponent of
$B(x)$. Since $\lambda(x_0)>0$, the system displays an exponential
dynamical instability and is hence called chaotic (in the sense of
Lyapunov) \cite{Rob95,Ott,ASY97,Beck}.

Writing down the analogue of Eq.~(\ref{eq:lyapbern}) for a given
differentiable map $F$, we get
\be
\Delta x_n = |x_n'-x_n|=|F^n(x_0')-F^n(x_0)|=: e^{n\lambda(x_0)} \Delta x_0\: (\Delta x_0\to0)\:,
\ee
which we can take as the definition of the Lyapunov exponent
$\lambda(x_0)$ that comes in as the exponential stretching rate on the
right hand side. Solving this equation for $\lambda(x_0)$ by using the
chain rule, it is not too hard to see \cite{ASY97} that this simple
procedure of calculating $\lambda$ can be generalized in terms of the
time (or Birkhoff) average
\be
\lambda(x)= \lim_{n\to\infty}\frac{1}{n} \sum_{i=0}^{n-1}
\ln\left|F'(x_i)\right| \label{eq:ljaptav}
\ee
with $x=x_0$. If the dynamical system defined by the map $F$ is
ergodic, the time average does not depend on the initial condition for
a typical $x$, $\lambda=\lambda(x)=const.$ It can be shown that the
Bernoulli shift is ergodic \cite{Do99}, and indeed, following
Eq.~(\ref{eq:ljaptav}), for $B$ we trivially find that $\lambda=\ln 2$
for all $x$. In particular, according to Birkhoff's theorem
\cite{Do99,KaHa95,ArAv68,TKS92}, for ergodic systems the time average is equal
to the ensemble average, which for the Lyapunov exponent of
one-dimensional maps reads
\be
\lambda=\langle\ln|F'(x)|\rangle_{\mu^*} := \int_J d\mu^* \ln|F'(x)| \label{eq:ljapens}\:.
\ee
Here $\mu^*$ is the invariant measure of the map. If the map exhibits
an SRB measure \cite{Young02,ER,Bal00}, we have
\be
d\mu^*=\rho^*(x)\:dx\:,\label{eq:musrb}
\ee
where $\rho^*(x)$ holds for the invariant density of the map. That is, the
measure $\mu^*$ has the nice property that it can be obtained by
integrating a density,
\be
\mu^*(A)=\int_Adx\:\rho^*(x)\:,\:A\subseteq J\: ,
\ee
which simplifies the calculation of the ensemble average
Eq.~(\ref{eq:ljapens}). For the Bernoulli shift it is not too
difficult to see \cite{Beck} that, for typical initial conditions, the
invariant density is $\rho^*(x)=1$. By combining
Eqs.~(\ref{eq:ljapens}) and (\ref{eq:musrb}), we get
\be
\lambda=\int_0^1 dx \rho^*(x) \ln 2 = \ln 2 \: .
\ee
This result is equal to the time average calculated above and confirms
the result obtained from our handwaving argument
Eq.~(\ref{eq:lyapbern}).

Lyapunov exponents are not the only quantities assessing the chaotic
character of a dynamical system. Pesin's Theorem
\cite{Do99,ER,Young02} states that for closed $C^2$ Anosov
\cite{ER,Do99} systems the Kolmogorov-Sinai (or metric) entropy
$h_{KS}$ is equal to the sum of positive Lyapunov exponents. For
one-dimensional maps that are expanding \cite{ASY97,Beck},
\be
\forall \: x \in J\quad |F'(x)|>1 \label{eq:expand}\:,
\ee 
this theorem boils down to
\be
\lambda=h_{KS}\:, \label{eq:pesin}
\ee
where \cite{Ott,Do99}
\be
h_{KS} := \lim_{n \rightarrow \infty}
-\frac{1}{n} \sum_{w\in \{W_i^n\}}\mu^*(w)\ln\mu^*(w)\:.\label{eq:hks}
\ee
Here $\mu^*(w)$ is the SRB measure of an element $w$ of the partition
$\{W_i^n\}$, and $n$ defines the level of refinement of the
partition. Note that in Eq.~(\ref{eq:hks}) we have assumed that the
partition is generating \cite{ER,KaHa95,BaPo97}. If $h_{KS}>0$ one
sometimes speaks of measure-theoretic chaos \cite{Beck}. For the
Bernoulli shift it is not too hard to calculate $h_{KS}$ from first
principles leading to $h_{KS}=\ln 2$ \cite{Ott,Beck}, which combined
with our previous result for the Lyapunov exponent is in line with
Pesin's theorem. This theorem can be formulated under weaker
assumptions, and it is believed to hold for a wider class of dynamical
systems than stated above. We remark that typically the KS-entropy is
much harder to calculate for a given dynamical system than Lyapunov
exponents. Hence, Pesin's theorem is often employed in the literature
for indirectly calculating the KS-entropy.

\section{Weak chaos and infinite ergodic theory}\label{sec:wciet}

Let us now consider a nonlinear generalization of our previous
piecewise linear model, which is known as the {\em Pomeau-Manneville
map} \cite{PoMa80},
\begin{figure}[t]
\centerline{\includegraphics[height=7cm]{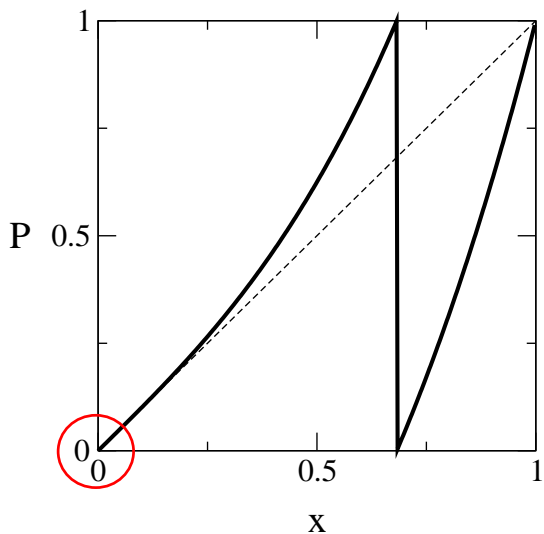}}
\caption{The Pomeau-Manneville map Eq.~(\ref{eq:pmmap}) for $a=1$ 
  and $z=3$.  Note that there is a marginal fixed point at $x=0$
  leading to the intermittent behavior depicted in
  Fig.~\ref{fig:ipoma}.}
\label{fig:poma} 
\end{figure}
\be
P_{a,z}(x) = x + ax^z \; \; \mbox{mod}\: 1\quad , \label{eq:pmmap}
\ee
see Fig.~\ref{fig:poma}, where following Eq.~(\ref{eq:eom}) the
dynamics is defined by $x_{n+1} = P_{a,z}(x_n)$. This map has the two
control parameters $a\ge 1$ and the exponent of nonlinearity $z\ge 1$.
For $a=1$ and $z=1$ the map reduces to the Bernoulli shift
Eq.~(\ref{eq:bern}), for $z>1$ it provides a nontrivial nonlinear
generalization of it. The nontriviality is due to the fact that in
this case the stability of the fixed point at $x=0$ becomes {\em
marginal} (sometimes also called indifferent, or neutral),
$P'_{a,z}(0)=1$. This implies that the map is non-hyperbolic, because
\cite{Dev89},
\begin{figure}[t]
  \centerline{\includegraphics[width=15cm]{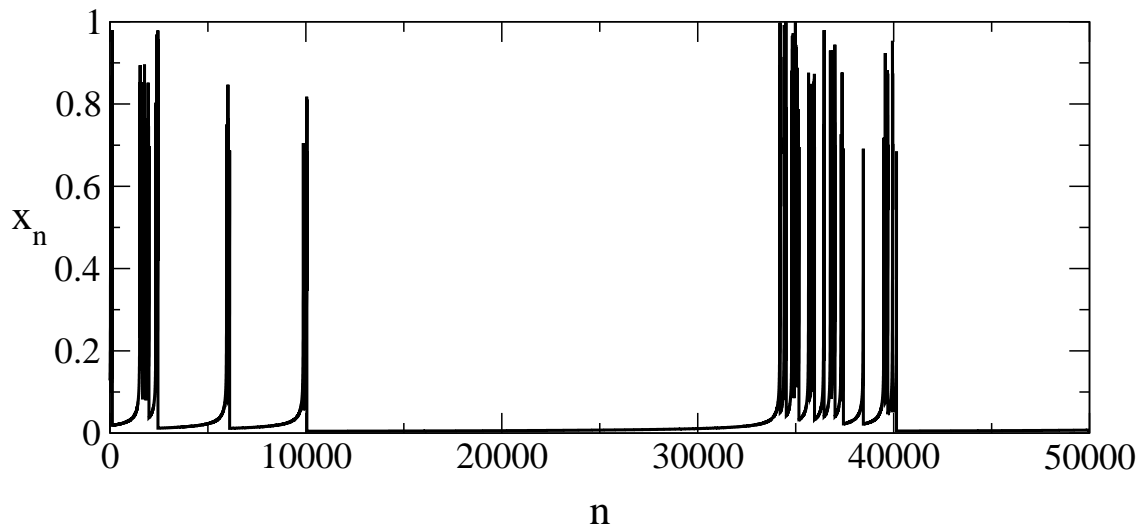}}
\caption{Phenomenology of intermittency in the Pomeau-Manneville map
  Fig.~\ref{fig:poma}: The plot shows the time series of position
  $x_n$ versus discrete time step $n$ for an orbit generated by the
  map Eq.~(\ref{eq:pmmap}), which starts at a typical initial condition
  $x_0$.}
\label{fig:ipoma} 
\end{figure}
\be
\not\exists N>0 \:\mbox{such that}\: \forall x \forall n\ge N \;|(P_{a,z}^n)'(x)|
\neq 1 \: ,
\ee
which is related to the fact that the map is not expanding anymore
according to Eq.~(\ref{eq:expand}). Since the map is smooth around
$x=0$, the dynamics resulting from the left branch of the map is
determined by the stability of this fixed point, whereas the right
branch is still of `Bernoulli shift-type' generating ordinary chaotic
dynamics. There is thus a competition in the dynamics between these
two different branches as illustrated by the time series displayed in
Fig.~\ref{fig:ipoma}: One can observe that long periodic `laminar
phases' determined by the marginal fixed point around $x=0$ are
interrupted by `chaotic bursts' reflecting the Bernoulli shift-type
part of the map with slope $a>1$ around $x=1$. This phenomenology is
the hallmark of what is called {\em intermittency} \cite{Schu,Ott}.

This seemingly small nonlinear modification of the Bernoulli shift has
dramatic consequences for the whole dynamics of the new system. We
discuss them step by step following our exposition of the Bernoulli
shift dynamics in Section \ref{sec:chsima}: The invariant density of
the Pomeau-Manneville map can be calculated to
\cite{Tha83,Zwei98,KoBa09,KoBa10,HoKl}
\be
\rho^*(x)\sim x^{1-z} \; (x\to 0)\:.
\ee
This singularity reflects the stickiness of trajectories to the
marginally unstable fixed point at $x=0$. Correspondingly, the measure
obtained by integrating this density is {\em non-normalizable} for
$z\ge2$ yielding the {\em infinite invariant measure}
\be
\mu^*(x)=\int_x^1 dy\: \rho^*(y)\to\infty\;(x\to0)\:.
\ee
The branch of erogdic theory exploring the ergodic properties of
infinite measure preserving dynamical systems is thus called {\em
infinite ergodic theory}; see Refs.~\cite{mtlnotes,Aar00,rzlnotes} for
introductions to this topic and \cite{Aar97} for an in-depth
mathematical treatment. The marginal fixed point has also an impact on
the dispersion of nearby trajectories, which can be calculated to
\cite{GaWa88,KoBa09,KoBa10,HoKl}
\be
\Delta x_n\sim\exp\left(n^{\frac{1}{z-1}}\right)\Delta x_0\;(z>2)\:.\label{eq:subexp}
\ee
In contrast to the Bernoulli shift, which according to
Eq.~(\ref{eq:lyapbern}) exhibits exponential sensitivity to intial
conditions, here we thus have a weaker {\em stretched exponential
sensitivity}. By repeating the calculation leading to
Eq.~(\ref{eq:ljaptav}), it is not hard to see that
Eq.~(\ref{eq:subexp}) yields a zero Lyapunov exponent,
\be
\lambda=0\:,
\ee
despite the fact that Fig.~\ref{fig:ipoma} displays irregular
dynamics. Dynamical systems where the separation of nearby
trajectories grows weaker than exponential, which implies that the
corresponding Lyapunov exponents are zero, have been coined {\em
weakly chaotic} \cite{ZaUs01,Gala03,vB04,ArCr05}. We remark, however,
that this denotation is not used unambiguously in the literature.
Most importantly, the standard definitions of Lyapunov exponents for
expanding and hyperbolic systems
Eqs.~(\ref{eq:ljaptav}),(\ref{eq:ljapens}) yield no good indicators of
irregular dynamics anymore, because they do not capture the
sub-exponential dispersion of trajectories. It is thus desirable to
come up with generalized definitions of ordinary chaos quantities,
which enable us to still assess this different type of chaotic
behavior by calculating quantities that yield non-zero values.

The way to achieve this goal is shown by advanced concepts of infinite
ergodic theory and corresponding generalized ergodic theorems. Recall
that Birkhoff's theorem implies that for observables which are
Lebesgue integrable, $f\in L^1$, we have
\cite{Do99,KaHa95,ArAv68,TKS92}
\be
\frac{1}{n}\sum_{i=0}^{n-1} f(x_i)=\langle f\rangle_{\mu^*} \:. \label{eq:birkh}
\ee
However, it turns out that for $z\ge 2$ the Birkhoff sum on the left
hand side does not converge anymore. Surprisingly, it becomes a random
variable that depends on initial conditions, and the equation breaks
down. This non-equivalence between time and ensemble averages became
known as {\em weak ergodicity breaking} in the physics literature,
see, e.g., Refs.~\cite{ReBa08,KRS08,SHB09} and further references
therein. It was observed experimentally in the anomalous statistics of
blinking quantum dots and plays also a crucial role for the anomalous
diffusion of atoms in optical lattices \cite{KRS08,ReBa08,SHB09}. Note
that physicists typically refer to ergodicity as the equality between
time and ensemble average, whereas mathematicians usually define
ergodicity via indecomposability \cite{ArAv68,KaHa95}.
Eq.~(\ref{eq:birkh}) then follows from this definition by using
Birkhoff's theorem. This should be kept in mind when referring to a
weak ergodicity breaking.

In case of $z\ge 2$ and $f\in L^1$ for our map, the nature of the
breakdown of Eq.~(\ref{eq:birkh}) is elucidated by the {\em
Aaronson-Darling-Kac theorem} \cite{Zwei00,ThZw06,rzlnotes}
\be
\frac{1}{a_n} \sum_{i=0}^{n-1} f(x_i) \stackrel{d}{\rightarrow} {\cal M}_{\alpha} \langle f\rangle_{\mu^*} \:(n \to \infty) \: ,
\ee
where the arrow holds for convergence in distribution. Here ${\cal
M}_{\alpha}$, $\alpha\in[0,1]$, denotes a non-negative real random
variable distributed according to the normalized Mittag-Leffler
distribution of order $\alpha$, which is characterized by its moments
\be
\langle {\cal M}_{\alpha}^r\rangle=r!\frac{(\Gamma(1+\alpha))^r}{\Gamma(1+r\alpha)}\:,\:r\ge0\:. \label{eq:adkthm}
\ee
For the Pomeau-Manneville map $P_{a,z}$ one can prove \cite{Zwei00}
that $a_n\sim n^{\alpha}$ with $\alpha:=1/(z-1)$. Integrating
Eq.~(\ref{eq:adkthm}) with respect to Lebesgue measure $m$ suggests
\be
\frac{1}{n^{\alpha}} \sum_{i=0}^{n-1} \langle f(x_i)\rangle_m\sim \langle f\rangle_{\mu^*}\:. \label{eq:adkint}
\ee
Note that for $z<2$ one has to choose $\alpha=1$, because the map
still exhibits an SRB measure, and Eq.~(\ref{eq:adkint}) becomes an
equality. However, for $z\ge 2$ we have an infinite invariant measure
that cannot be normalized, hence here Eq.~(\ref{eq:adkint}) remains a
proportionality, unless we fix this constant by other constraints.

These known facts from infinite ergodic theory motivate to suitably
define generalized chaos quantities, which assess weakly chaotic
dynamics by yielding non-zero values. Following the left hand side of
Eq.~(\ref{eq:adkint}), by choosing $f(x)=\ln|P'_{a,z}(x)|$ we define
the {\em generalized Lyapunov exponent} as
\be
\Lambda :=  \lim_{n\to\infty} 
\frac{\Gamma(1+\alpha)}{n^{\alpha}} \sum_{i=0}^{n-1} 
\langle\ln \left| M'(x_i) \right|\rangle_m\>\label{eq:defgle}\:.
\ee
The inclusion of the gamma function in the numerator is not obvious at
this point, however, it turns out to be convenient when calculating
$\Lambda$ for the Pomeau-Manneville map
\cite{HoKl}. Interestingly, it is precisely the same canonical choice as is
made in other areas of anomalous dynamics \cite{KKCSG07}. Analogously,
we amend Eq.~(\ref{eq:hks}) to define the {\em generalized KS entropy}
as
\be
H_{KS} := \lim_{n\to\infty}
-\frac{\Gamma(1+\alpha)}{n^{\alpha}} \sum_{w\in \{W_i^n\}} \mu^*(w) \ln
\mu^*(w)\:.\label{eq:ghks}
\ee
Both quantities can be calculated independently for the piecewise
linearization of $P_{a,z}$ proposed in Ref.~\cite{GaWa88} by applying
the thermodynamic formalism \cite{Rue78,Beck} in combination with
transfer operator methods
\cite{PrSl92,TaGa02,TG04}. As a result, one obtains \cite{HoKl}
\be
H_{KS}=\Lambda\:,\label{eq:gpesin}
\ee
which may be considered as a generalization of Pesin's formula
Eq.~(\ref{eq:pesin}) to anomalous dynamics. Related generalizations of
chaos quantities, and other versions of a generalized Pesin formula,
have been discussed in Refs.~\cite{KoBa09,KoBa10,AkAi10,SaVe12}. We
remark, however, that in the mathematical literature there is the
well-known formula by Rokhlin \cite{Kell98}, which for the
Pomeau-Manneville map reads \cite{Tha83,Zwei00}
\begin{figure}[t]
  \centerline{\includegraphics[height=6cm]{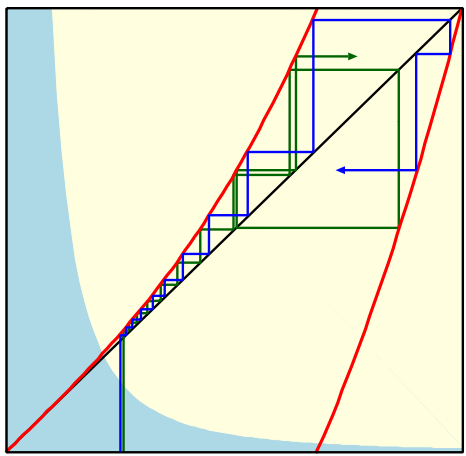}}
\caption{Illustration of the interplay between weak chaos, infinite measures, 
and anomalous dynamics in the Pomeau-Manneville map
Eq.~(\ref{eq:pmmap}) shown by the thick red lines: Anomalous dynamics
is indicated by the irregular behaviour of the two single
trajectories. Weak chaos is exemplified by the divergence of the two
trajectories starting at nearby initial conditions. The light blue
area in the lower left corner depicts the shape of the infinite
invariant density, which diverges at the marginal fixed point of the
map \cite{wchaos11}.}
\label{fig:wchaoslogo} 
\end{figure}
\be
h_{Kr}=\langle\ln|P'_{a,z}(x)|\rangle_{\mu^*}\:.
\ee
Here the left hand side holds for the so-called Krengel entropy.  In
case of finite invariant measures one can show that $h_{KS}=h_{Kr}$,
the right hand is the Lyapunov exponent defined via the ensemble
average Eq.~(\ref{eq:ljapens}), and Rokhlin's formula boils down to
Pesin's formula Eq.~(\ref{eq:pesin}). For infinite invariant measures,
one can show that $h_{Kr}=H_{KS}$ \cite{HoKl}. Combining Rokhlin's
formula with the integrated form of the Aaronson-Darling-Kac theorem
Eq.~(\ref{eq:adkint}) by using $f(x)=\ln|P'_{a,z}(x)|$, exploiting the
definition Eq.~(\ref{eq:defgle}) for the generalized Lyapunov
exponent, and by fixing the constant of proportionality in
Eq.~(\ref{eq:adkint}) with respect to Lebesgue initial measure, one
recovers Eq.~(\ref{eq:gpesin}). One may thus argue that, within this
setting, Rokhlin's formula is a generalization of Pesin's formula for
infinite measure-preserving transformations, and that
Eq.~(\ref{eq:gpesin}) is an illustration of it, worked out for the
example of the Pomeau-Manneville map \cite{HoKl}.

The main theoretical objects of discussion in this subsection are
shown together in Fig.~\ref{fig:wchaoslogo}. This figure actually
represents the logo of the conference about {\em Weak chaos, infinite
ergodic theory and anomalous dynamics} that was referred to in the
introduction
\cite{wchaos11}, from which the title of this book chapter derives.

\section{A generalized hierarchy of chaos}

We conclude this section by embedding the previous results into the
more general context of irregular deterministic dynamics
\cite{Kla06}. There exist in fact further fundamental types of
dynamics that are intermediate between strongly chaotic, in the sense
of exponential sensitivity quantified by a positive Lyapunov exponent,
and trivially being non-chaotic in terms of purely regular
dynamics. These different types of irregular dynamics can be
characterized by suggesting a classification of chaotic behavior based
on the dispersion of initially infinitesimally close trajectories.

\begin{figure}[t]
  \centerline{\includegraphics[height=15cm,angle=-90]{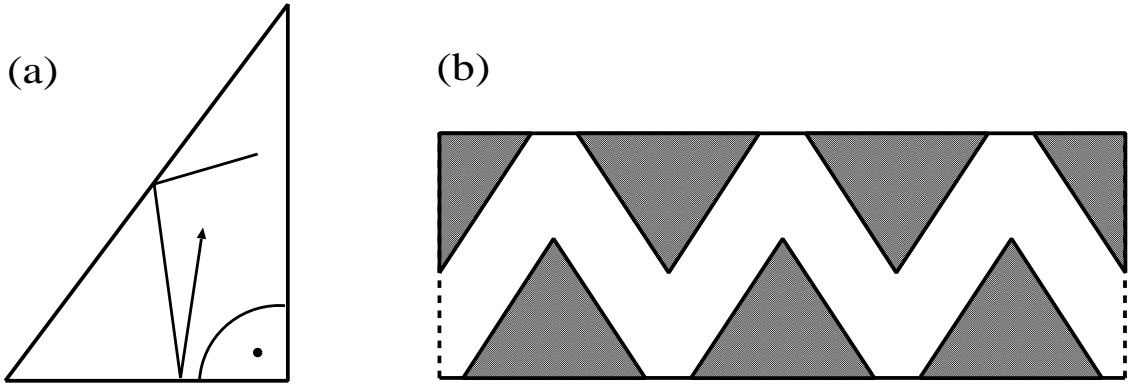}}
\caption{Two simple examples of two-dimensional polygonal billiards 
\cite{Kla06}: A 
particle with unit velocity moves inside the depicted geometric
domains by scattering elastically with their boundaries. (a) shows a
right triangular billiard \cite{ArCaGu97}, (b) the triangle channel,
where a unit cell with triangular scatterers is spatially continued
along the line \cite{LWH02}.}
\label{fig:poly}
\end{figure}

We start from the general expression for the asymptotic growth of the
displacement $\Delta(t)$ of two trajectories generated by dynamics in
continuous time $t$ in the form of \cite{GaWa88}
\be
\ln\Delta(t)\sim t^{\nu_0}(\ln t)^{\nu_1} \: , \: 0\le \nu_0 \: , \: \nu_1 \in \mathbb{R} \quad . \label{eq:chcl}
\ee
If $\nu_0=1\:,\:\nu_1=0$ we recover the usual exponential dynamical
instability of Eq.~(\ref{eq:lyapbern}),
\be
\Delta(t)\sim \exp(\lambda t)\:,
\ee
representing {\em Lyapunov chaos} \cite{Ott}, whose strength is well
quantified by the maximal positive Lyapunov exponent $\lambda$. As
discussed before, if $\Delta(t)$ grows weaker than exponential, one
speaks of {\em weak chaos} \cite{ZaUs01,Gala03,vB04,ArCr05}. The
regime of Eq.~(\ref{eq:chcl}) with $0<\nu_0<1$ or $\nu_0=1$ and
$\nu_1<0$, which is typical for intermittent dynamics, was
characterized as {\em sporadic} by Gaspard and Wang \cite{GaWa88};
cf.\ Eq.~(\ref{eq:subexp}) and our respective discussion, as well as
further details of this dynamics as presented in the following
section. Here the dynamical instability is either of stretched
exponential type or it is exponential with logarithmic corrections,
\be
\Delta(t)\sim \exp(t^{\nu_0}(\ln t)^{\nu_1}) \: .
\ee
Eq.~(\ref{eq:chcl}) with $\nu_0=0$ and $\nu_1=1$, on the other hand,
yields purely algebraic dispersion,
\be
\Delta(t)\sim t^{\nu_2} \:,\:0<\nu_2\: ,
\ee
for which Zaslavsky and Edelman \cite{Zas02,ZaEd03} suggested the term
{\em pseudochaos}.\footnote{We remark that in
Refs.~\cite{Zas02,ZaEd03,ZCLGE05} one finds several slightly different
definitions of pseudochaos. Here we refer to the first one stated in
Ref.~\cite{Zas02}.}  Note that algebraic dispersion with logarithmic
corrections may also exist,
\be
\Delta(t)\sim t^{\nu_2}(\ln t)^{\nu_3} \:,\nu_3\in\mathbb{R}\quad ,
\ee
covering a slightly larger class of dynamical systems. A prominent
class of dynamical systems exhibiting algebraic dispersion are
polygonal billiards; two examples are depicted in Fig.~\ref{fig:poly}.
They represent the special case of pseudochaotic dynamics with
$\nu_2=1$ for which the dispersion is strictly linear in time,
\be
\Delta(t)\sim t \quad .
\ee
However, in contrast to Lyapunov chaos and our weakly chaotic
generalizations, here the linear dispersion does not actually capture
the essential mechanism leading to dynamical randomness. For example,
according to this classification both free flights and polygonal
billiards of genus one, which clearly exhibit regular dynamics, are
also pseudochaotic. As is discussed for the example of rational
billiards, e.g., in Ref.~\cite{Kla06}, in polygonal billiards
complicated topologies reflecting the existence of corners, which
yield pseudohyperbolic fixed points and pseudointegrability, provide
the source of non-trivial irregular dynamics. One is thus tempted to
speak of {\em topology-induced chaos}\footnote{This should not be
confused with {\em topological chaos} as defined in Ref.~\cite{Gasp}.} 
as a subclass of pseudochaos if there is linear dispersion on surfaces
that are not integrable. {\em Pseudointegrable} rational billiards
then form a subclass of topology-induced chaotic dynamical
systems. Surprisingly, systems with linear dispersion generating
non-trivial dynamics due to complicated topological structures may
still exhibit ergodic and transport properties as they are usually
associated with Lyapunov unstable chaotic dynamical systems. The
trivial end point of this attempt of a generalized classification of
chaotic dynamics on the basis of dispersion is simply the purely
regular, or periodic, case of $\Delta(t)=const.$

\chapter{Anomalous diffusion}\label{sec:anodif}

We now establish a cross-link between weakly chaotic dynamics as
discussed in the previous section and the problem of deterministic
diffusion. The main question we address is what type of diffusion
arises if we suitably spatially extend a simple dynamical system
exhibiting anomalous dynamics. We first set up our model, which can be
considered as a purely deterministic, anomalous version of a random
walk on the line, and introduce the concept of anomalous diffusion. We
then outline continuous time random walk theory, a standard tool in
the theory of stochastic processes to study anomalous diffusion. The
results obtained from this theory, worked out for our model, are
compared to results from computer simulations. We conclude this
chapter by deriving on the basis of this theory a generalized,
fractional diffusion equation that approximately reproduces the
probability density function (PDF) of our model.

\section{A simple model generating anomalous diffusion}

\begin{figure}[t]
\centerline{\includegraphics[width=9.7cm]{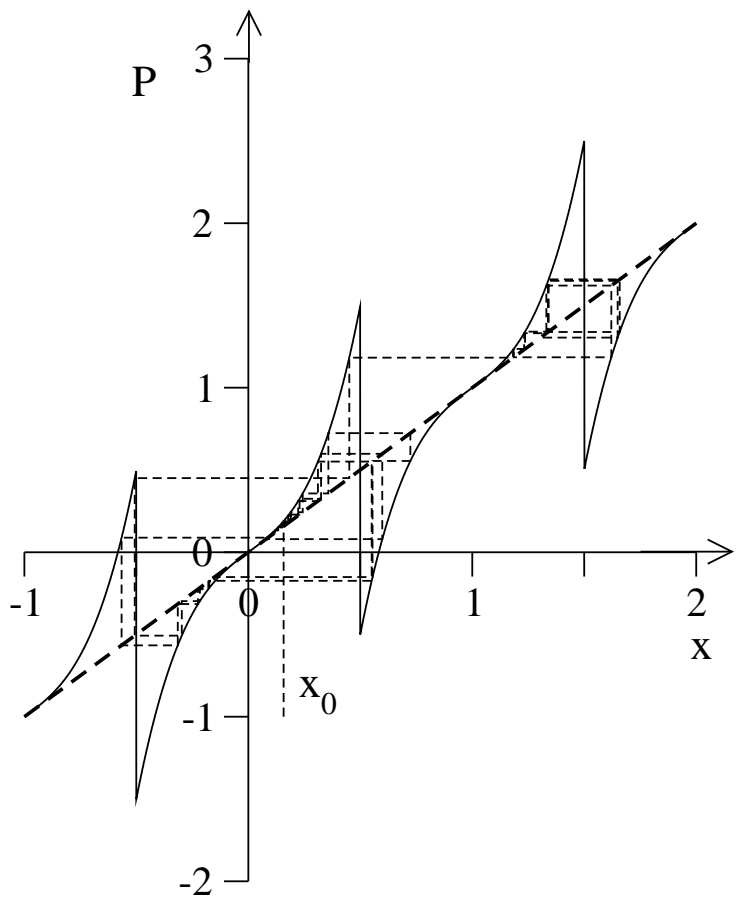}}
\caption{The Pomeau-Manneville map Fig.~\ref{fig:poma}, Eq.~(\ref{eq:pmmap}), 
lifted symmetrically onto the whole real line such that it generates
subdiffusion.}
\label{fig:liftedpm} 
\end{figure}

A straightforward way to define a spatially extended dynamical system
based on the Pomeau-Manneville map discussed in
Section~\ref{sec:wciet} is as follows: By using 
\be
P_{a,z}(x) = x +a x^z\;,\; 0 \le x <\frac{1}{2}\label{eq:pomadif}
\ee
of Eq.~(\ref{eq:pmmap}) without the modulus, as well as reflection
symmetry,
\be
P_{a,z}(-x)=-P_{a,z}(x)\:,\label{eq:refl}
\ee
we continue this map onto the whole real line by a {\em lift of degree
one} \cite{GF2,GeNi82,SFK}
\be
P_{a,z}(x + 1) = P_{a,z} (x) + 1\:.
\ee
The resulting model \cite{GeTo84,ZuKl93a} is displayed in
Fig.~\ref{fig:liftedpm}. Here points are not restricted anymore onto
the unit interval. Due to the coupling between different unit cells by
eliminating the modulus, there are now `jumps' possible from unit
interval to unit interval. One may thus think of this dynamical system
as a fully deterministic, anomalous version of a simple random walk on
the line. A basic question is now which type of diffusion is generated
by this model? As usual, the diffusive behavior is quantified by the
mean square displacement (MSD) defined by
\be
\langle x^2\rangle :=\int dx\;x^2\rho_n(x) \label{eq:vari}\:,
\ee
where $\langle x^2\rangle$ is the second moment of the position PDF
$\rho_n(x)$ at time step $n$. Starting from a given initial PDF
$\rho_0(x)$ at time step $n=0$, points, or point particles, will
spread out over the whole real line, as quantified by
$\rho_n(x)$. Surprisingly, by calculating this MSD both analytically
and from computer simulations one finds
\cite{GeTo84,ZuKl93a} that for $z>2$
\be
\left< x^2 \right>\sim\:n^{\alpha}\quad , \quad \alpha<1\quad(n\to\infty)\: . \label{eq:amsd}
\ee
If one defines the diffusion coefficient of the system in the standard way by
\be
D:=\lim_{n\to\infty}\frac{\langle x^2\rangle}{2n}\:,
\ee
Eq.~(\ref{eq:amsd}) implies that $D=0$, despite the fact that
particles can go anywhere on the real line as illustrated in
Fig.~\ref{fig:liftedpm}. While a process like Brownian motion leads to
$D>0$, here we thus encounter a very different type of diffusion: If
the exponent $\alpha$ in the temporal spreading of the MSD
Eq.~(\ref{eq:amsd}) of an ensemble of particles is not
equal to one, one speaks of {\em anomalous diffusion}
\cite{MeKl00,KRS08}. If $\alpha<1$ one says that there is {\em
subdiffusion}, for $\alpha>1$ there is {\em superdiffusion}. In case
of linear spreading with $\alpha=1$ one refers to {\em normal
diffusion}. The constant
\be
K:=\lim_{n\to\infty}\frac{<x^2>}{n^{\alpha}}\quad , \label{GDC_Def}
\ee
where in case of normal diffusion in one dimension $K=2D$, is called
the {\em generalized diffusion coefficient}.\footnote{In detail, the
definition of a generalized diffusion coefficient is a bit more subtle
\cite{KKCSG07}.} For our simple map model depicted in 
Fig.~\ref{fig:liftedpm} we will first calculate the MSD analytically
by means of stochstic {\em continuous time random walk} (CTRW)
theory. By comparing the analytical formula with results from computer
simulations, we will then focus on how $K$ behaves as a function of
$z$ for fixed $a$ revealing some interesting, non-trivial properties.

\section{Continuous time random walk theory}

Pioneered by Montroll, Weiss and Scher \cite{MW65,MS73,SM75}, CTRW
theory yields perhaps the most fundamental theoretical approach to
explain anomalous diffusion \cite{BoGe90,Weiss94,EbSo05}. In further
groundbreaking works by Geisel et al.\ and Klafter et al., this method
was then adapted to sub- and superdiffusive deterministic maps
\cite{GeTo84,GNZ85,ShlKl85,ZuKl93a}

The basic assumption of this approach is that diffusion can be
decomposed into two stochastic processes characterized by waiting
times and jumps, respectively. Thus one has two sequences of
independent identically distributed random variables, namely a
sequence of positive random waiting times $T_1, T_2, T_3,
\ldots$ with PDF $w(t)$ and a sequence of random jumps $\zeta_1, \zeta_2,
\zeta_3, \ldots$ with a PDF $\lambda(x)$. For
example, if a particle starts at point $x = 0$ at time $t_0 = 0$ and
makes a jump of length $\zeta_n$ at time $t_n = T_1 + T_2 + ... +
T_n$, its position is $x = 0$ for $0 \le t < T_1 = t_1$ and $x =
\zeta_1 + \zeta_2 + ...  + \zeta_n$ for $t_n \le t < t_{n+1}$. The
probability that at least one jump is performed within the time
interval $[0,t)$ is then $\int_0^t dt' w(t')$ while the probability
for no jump during this time interval reads $\Psi(t) = 1 -
\int_{0}^{t} dt' w(t')$. The master equation for the PDF
$P(x,t)$ to find a particle at position $x$ and time
$t$ is then
\begin{equation}
  \label{master_eq} P(x,t) = \int_{-\infty}^{\infty} dx' \lambda (x
  - x') \int_{0}^{t} dt' \; w(t -  t') \; P(x',t') + \Psi(t) \delta(x)\: ,
\end{equation}
which has the following probabilistic meaning: The PDF
to find a particle at position $x$ at time $t$ is equal to
the PDF to find it at point $x'$ at some
previous time $t'$ multiplied with the transition probability to get
from $(x',t')$ to $(x,t)$ integrated over all possible values of $x'$
and $t'$. The second term accounts for the probability of remaining at
the initial position $x=0$. The most convenient representation of this
equation is in Fourier-Laplace space,
\begin{equation} \label{times_PDF} 
\hat{\tilde{P}} (k,s) =
\int_{-\infty}^{\infty} dx \; e^{i k x} \int_{0}^{\infty} dt \;
e^{-st} P(x,t) \quad , 
\end{equation}
where the hat stands for the Fourier transform and the tilde for the
Laplace transform. Respectively, this function obeys the
Fourier-Laplace transform of Eq.~(\ref{master_eq}), which is called
the Montroll-Weiss equation \cite{MW65,MS73,SM75},
\index{Montroll-Weiss equation|ibold}
\begin{equation} \label{Montroll_Weiss} 
\hat{\tilde{P}} (k,s) =
\frac{1 - \tilde{w}(s)}{s} \frac{1}{1-
\hat{\lambda}(k)\tilde{w}(s)}\quad .  \end{equation}
The Laplace transform of the MSD can be obtained by differentiating
the Fourier-Laplace transform of the PDF,
\begin{equation}
\label{Laplace_MSD}
\tilde{ \left< x^2 (s) \right> } = \int_{-\infty}^{\infty} dx \; x^2 \tilde{P}(x,s) =
\left. - \frac{\partial^2 \hat{\tilde{P}} (k,s) }{\partial k^2} \right|_{k=0}\quad .
\end{equation}
In order to calculate the MSD within this theory, it thus suffices to
know $\lambda(x)$ and $w(t)$ generating the stochastic process. For
one-dimensional maps of the type of
Eqs.~(\ref{eq:pomadif}),(\ref{eq:refl}), by exploiting the symmetry of
the map the waiting time distribution can be calculated from the
approximation
\begin{equation}
\label{cont_time_map}
x_{n+1} - x_n \simeq \frac{dx_t}{dt} = a x_t^z, \; \; x \ll 1 \quad ,
\end{equation}
where we have introduced the continuous time $t\ge 0$. This equation
can be solved for $x_t$ with respect to an initial condition
$x_0$. Now one needs to define when a particle makes a ``jump'', as
will be discussed below. By inverting the solution for $x_t$, one can
then calculate the time $t$ a particle has to wait before it makes a
jump as a function of the initial condition $x_0$. This information
determines the relation between the waiting time PDF $w(t)$ and the as
yet unknown PDF of injection points,
\begin{equation}
\label{cont_time_map_sol4}
w(t) \simeq P_{in}(x_0) \left| \frac{dx_0}{dt}\right| \quad .
\end{equation}
Making the assumption that the PDF of injection points
is uniform, $P_{in} \simeq 1$, the waiting time PDF is
straightforwardly calculated from the knowledge of $t(x_0)$.
The second ingredient that is needed for the CTRW approach is the jump
PDF. Standard CTRW theory takes jumps between
neighbouring cells {\em only} into account leading to the ansatz
\cite{GeTo84,ZuKl93a}
\begin{equation}
\label{jump_PDF}
\lambda (x) = \delta (|x| - 1) \quad .
\end{equation}
It turns out that in order to qualitiatively reproduce the dependence
of the generalized diffusion coefficient $K=K(z,a)$
Eq.~(\ref{GDC_Def}) on the map's two control parameters $z$ and $a$,
one needs to modify the standard theory at three points
\cite{KCKSG06,KKCSG07}: Firstly, the waiting time PDF 
must be calculated according to the unit interval $[0,1]$, not
according to $[-0.5,0.5]$, which is an alternative but not appropriate
choice \cite{RKdiss,dcrc}, yielding
\begin{equation}
 w(t)=a\left( 1+a(z-1)t\right) ^{-\frac z{z-1}}\: .
\label{waiting_time_pdf}
\end{equation}
However, this PDF also accounts for {\em
attempted} jumps to another cell, since after a step the particle may
stay in the same cell with a probability of $(1 - p)$.  The latter
quantity is roughly determined by the size of the escape region $p = (
1 - 2 x_c )$ with $x_c$ as a solution of the equation $x_c + a
x^{z}_{c} = 1$. We thus model this fact, secondly, by a jump length
distribution in the form of
\begin{equation}
\label{real_jump_PDF}
\lambda (x) = \frac{p}{2} \delta (\left| x \right| - l) + (1 - p) \delta
(x)\:.
\end{equation}
Thirdly, in order to capture the dependence of $K$ on $z$ for fixed $a$,
we define a typical jump length as
\begin{equation}
\label{int_jump_length}
l = \left\{ | [M_{a,z}(x)] | \right\} \: ,
\end{equation}
where the square brackets stand for the floor function, which gives
the coarse-grained displacement in units of elementary cells. The
curly brackets denote both a time and ensemble average over points
leaving a box. Note that for capturing the dependence of $K$ on $a$
for fixed $z$ a different definition of the jump length is appropriate
\cite{KCKSG06,KKCSG07}. Working out the modified CTRW approximation sketched
above by taking these three details into account one obtains the
result for the exponent $\alpha$ of the MSD, Eq.~(\ref{eq:amsd}),
\begin{equation}
\alpha = \begin{cases} 1 , & 1 \le z < 2 \cr \label{eq:alpha}
\frac{1}{z-1}, & 2\le z  \end{cases} \:,
\end{equation} 
which matches to the standard theory \cite{GeTo84,ZuKl93a}. This
result is in excellent agreement to simulations for a broad range of
control parameters $z$ and $a$. Building on this result, the
generalized diffusion coefficient can be calculated to
\begin{figure}[t]
\centerline{\includegraphics[height=10cm,angle=-90]{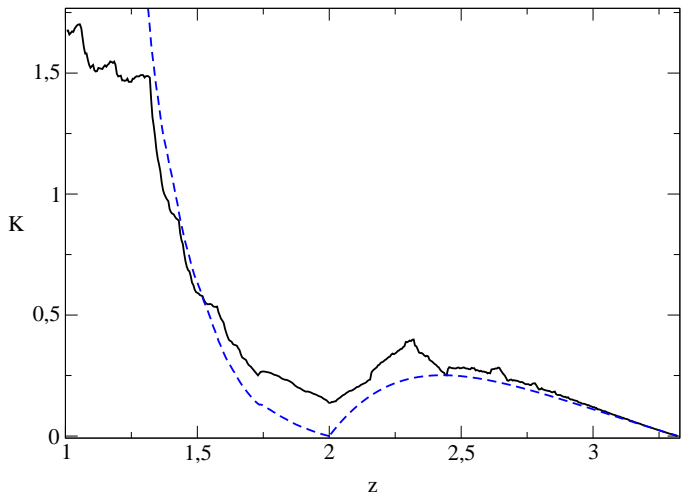}}
\caption{The generalized diffusion coefficient $K$, Eq.~(\ref{GDC_Def}), 
for the spatially extended Pomeau-Manneville map displayed in
Fig.~\ref{fig:liftedpm} as a function of $z$ for $a=5$. The bold black
line depicts computer simulation results. The dashed line corresponds
to the modified CTRW approximation
Eqs.~(\ref{int_jump_length}),(\ref{GDC_CTRW}) \cite{KKCSG07}.}
\label{fig:gdckz} 
\end{figure}
\begin{equation}
K  = p l^2\times\left\{
\begin{array}{l@{\quad,\quad}l}
 a^{\gamma} \sin (\pi \gamma) / \pi \gamma^{1+\gamma} & 0 < \gamma < 1 \\
 a (1-1/\gamma) & 1 \le \gamma < \infty
\end{array}\right. \quad ,\label{GDC_CTRW}
\end{equation}
with $\gamma:=1/(z-1)$, which for $z\ge 2$ is identical to $\alpha$ of
Eq.~(\ref{eq:alpha}). In Fig.~\ref{fig:gdckz} this analytical
approximation for $K$ is compared with computer simulation results as
a function of $z$ for fixed $a$. There is good qualitative agreement
between theory and simulations for $z>2$, which converge
asymptotically to each other for large $z$. For $z<2$ there is
reasonable qualitative agreement, though quantitative deviations, for
$z$ close to $2$ while the theory does not work anymore for $z\to0$, a
problem that is discussed in Ref.~\cite{KKCSG07}.

Remarkably, the $K(z)$ obtained from simulations does not appear to be
a smooth function of the control parameter, which is at variance with
the prediction of CTRW theory Eq.~(\ref{GDC_CTRW}). This non-smooth
parameter dependence is not due to numerical errors (which here are
very difficult to assess, as discussed below) but a well-known
phenomenon for this type of systems. It has first been discovered for
the normal diffusive case of this map with $z=1$, where the diffusion
coefficient $D=K/2$ has been studied both numerically and analytically
as a function of the slope $a$ as a control parameter
\cite{RKD,KlDo99}. Note that for $z=1$ the Pomeau-Manneville map boils
down to a parameter-dependent, generalized version of the Bernoulli
shift Eq.~(\ref{eq:bern}).

We do not wish to further elaborate on the fractal parameter
dependencies of transport coefficients in simple deterministic
dynamical systems, an interesting phenomenon that has been discussed
in detail in Refs.~\cite{RKdiss,Kla06}. Rather, we focus on the
behaviour of the generalized diffusion coefficient at the point
$z=2$. According to the exponent $\alpha$ of the MSD given by
Eq.~(\ref{eq:alpha}), here the map exhibits a transition from normal
to anomalous diffusion, which one may classify as a dynamical phase
transition \cite{Wang89b,Beck}. As can be seen in
Fig.~\ref{fig:gdckz}, right around this transition point there are
significant deviations between CTRW theory and the simulation
results. Most notably, at $z=2$ the CTRW approximation forms a
non-differentiable little wedge by predicting $K(2)=0$, whereas the
simulations yield $K(2)>0$. By increasing the computation time one
indeed finds very slow convergence of the simulation data towards the
CTRW solution \cite{KKCSG07}.

The explanation of these deviations, and of the phenomenon of a
complete suppression of the generalized diffusion coefficient right at
the transition point, is obtained by carrying out a refined analysis
by means of CTRW theory. For a long time it was known already that at
$z=2$, the MSD behaves like $\langle x^2\rangle\sim n/\ln
n\:(n\to\infty)$ \cite{GeTo84,ZuKl93a}. Note that according to our
definition of the generalized diffusion coefficient
Eq.~(\ref{GDC_Def}) this logarithmic depencence was incorporated into
the strength of the diffusion coefficient, otherwise our analytical
CTRW approximation would not have been continuous at $z=2$. By taking
into account higher-order terms when performing the CTRW theory
calculations, which correspond to lower-order terms in time for the
MSD, one arrives at
\be
\langle x^2\rangle \sim
\begin{cases}
\frac{n}{\ln n} \:,\: n<n_{cr} \:\mbox{and}\: \sim n \:,\: 
n\gg n_{cr} \:,\: & z<2\\
\frac{n}{\ln n} \:,\: & z=2\\
\frac{n^{\alpha}}{\ln n} \:,\: n<\tilde{n}_{cr} \:\mbox{and}\: 
\sim n^{\alpha} \:,\: n\gg\tilde{n}_{cr} \:,\: & z>2\label{eq:logcorr}\:.
\end{cases}
\ee
Here $n_{cr}$ and $\tilde{n}_{cr}$ are crossover times that can be
calculated exactly in terms of the map's control parameters. For
$z\to2$ both these crossover times diverge, and one arrives at the
asymptotic $n/\ln n$ dependence mentioned before. The perhaps
surprising fact is that around the transition point these
multiplicative logarithmic corrections still survive for long but
finite time, in agreement with computer simulation results. In other
words, these logarithmic corrections lead to an ultraslow convergence
of the simulation results thus explaining the deviations between
long-time CTRW theory and simulations shown in
Fig.~\ref{fig:gdckz}. But more importantly, these logarithmic terms
dominate the strength of the generalized diffusion coefficient around
the transition point from normal to anomalous diffusion eventually
yielding a full suppression of this quantity right at the transition
point. It can be conjectured that the presence of such multiplicative
logarithmic corrections around transition points between normal and
anomalous diffusion is a typical scenario in this type of systems
\cite{KKCSG07}.

\section{A fractional diffusion equation}

We now turn to the PDFs generated by the
lifted map Eq.~(\ref{eq:pmmap}). As we will show now, CTRW theory not
only predicts the power $\alpha$ correctly but also the form of the
coarse grained PDF $P(x,t)$ of displacements. Correspondingly the
anomalous diffusion process generated by our model is not described by
an ordinary diffusion equation but by a generalization of it. Starting
from the Montroll-Weiss equation and making use of the expressions for
the jump and waiting time distributions Eqs.~(\ref{jump_PDF}),
(\ref{waiting_time_pdf}), we rewrite Eq.~(\ref{Montroll_Weiss}) in the
long-time and -space asymptotic form
\begin{equation} 
s^{\gamma}\hat{\tilde{P}} - s^{\gamma - 1} = - \frac{p l_i^2}{2 c b^{\gamma}} k^2
\hat{\tilde{P}} \label{Montroll-Weiss_2} 
\end{equation} 
with $c = \Gamma (1-\gamma)$ and $b=\gamma /a$.  For the initial
condition $P(x,0)=\delta (x)$ of the PDF we have
$\hat{P}(k,0)=1$. Interestingly, the left hand side of this equation
corresponds to the definition of the {\em Caputo fractional
derivative} of a function $G$ \cite{Podl99,Mai97},
\begin{equation} \frac{\partial^{\gamma}
G}{\partial t^{\gamma}} := \frac{1}{\Gamma (1-\gamma)} \; \int_0^t
dt^{^{\prime }}(t-t^{^{\prime }})^{-\gamma }\frac{\partial G}{\partial
t^{^{\prime }}} \label{Caputo} \quad ,
\end{equation}
in Laplace space,
\begin{equation}
\int_{0}^{\infty} dt \; e^{-st} \frac{\partial^{\gamma} G}{\partial
t^{\gamma}} = s^{\gamma} \tilde{G} (s) - s^{\gamma - 1} G(0) \quad .
\label{Caputo_Laplace} 
\end{equation}
Thus, fractional derivatives come naturally into play as a suitable
mathematical formalism whenever there are power law memory kernels in
space and/or time generating anomalous dynamics; see, e.g.,
Refs.~\cite{SKB02,MeKl00} for short introductions to fractional
derivatives and Ref.~\cite{Podl99} for a detailed exposition. Turning
back now to real space, we thus arrive at the time-fractional
diffusion equation
\begin{equation}
\frac{\partial ^\gamma P(x,t)}{\partial t^\gamma } = D \; \frac{\partial ^2P}{\partial x^2}
\label{Frac_Dif_Eq}
\end{equation}
with $D=K\Gamma (1 + \gamma)/2$, $0<\gamma <1$, which is an example of
a {\em fractional diffusion equation} generating subdiffusion. Note
the existence of the gamma function in the numerator defining $D$,
which is analogous to the appearance of the gamma function in our
generalized chaos quantities
Eqs.~(\ref{eq:defgle}),(\ref{eq:ghks}). For $\gamma=1$ we recover the
ordinary diffusion equation. The solution of Eq.~(\ref{Frac_Dif_Eq})
can be expressed in terms of an M-function of Wright type \cite{Mai97}
and reads
\begin{figure}[t]
\centerline{\includegraphics[width=9.5cm]{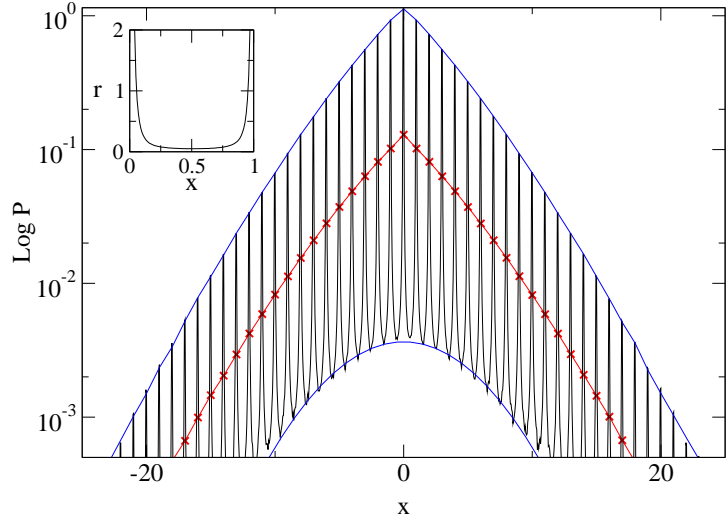}}
\caption{Comparison of the probability density obtained from simulations of 
  the lifted map Eq.~(\ref{eq:pmmap}) (oscillatory structure) with the
  analytical solution Eq.\ (\ref{ctrw_sol}) of the fractional
  diffusion equation Eq.~(\ref{Frac_Dif_Eq}) (continuous line in the
  middle) for $z=3$ and $a=8$. The probability density was computed
  from $10^7$ particles after $n=10^3$ iterations. For the generalized
  diffusion coefficient in Eq.~(\ref{ctrw_sol}) the simulation result
  was used. The crosses (x) represent the numerical results coarse
  grained over unit intervals. The upper and the lower curves
  correspond to fits with a stretched exponential and a Gaussian
  distribution, respectively. The inset depicts the probability
  density function for the map on the unit interval with periodic
  boundaries.}
\label{fig:fctrw4}
\end{figure}
\begin{equation}
P(x,t)=\frac 1{2\sqrt{D}t^{\gamma /2}}M\left( \xi,\frac \gamma
2\right) \quad .
\label{ctrw_sol}
\end{equation}
Fig.~\ref{fig:fctrw4} demonstrates an excellent agreement between the
analytical solution Eq.\ (\ref{ctrw_sol}) and the PDF obtained from
simulations of the map Eq.~(\ref{eq:pmmap}) if it is coarse grained
over unit intervals.  However, it also shows that the coarse graining
eliminates a periodic fine structure that is not captured by
Eq.~(\ref{ctrw_sol}). This fine structure derives from the
`microscopic' invariant density of an elementary cell (with periodic
boundaries) as shown in the inset of Fig.~\ref{fig:fctrw4}
\cite{RKdiss}. The singularities are due to the marginal fixed points
of the map, where points are trapped for long times.  Remarkably, that
way the microscopic origin of the intermittent dynamics is reflected
in the shape of the PDF on the whole real line: From
Fig.~\ref{fig:fctrw4} it is seen that the oscillations in the density
are bounded by two functions, the upper curve being of a stretched
exponential type while the lower is Gaussian. These two envelopes
correspond to the laminar and chaotic parts of the motion,
respectively.

\chapter{Anomalous fluctuation relations}\label{sec:afrc}

After having accomplished a transition from deterministic dynamics to
stochastic modeling in the previous section, for the reminder of this
chapter we fully focus on stochastic systems. First, we discuss a
remarkable finding in nonequilibrium statistical mechanics that was
widely investigated over the past two decades, which are fluctuation
relations. After providing a brief outline of what they are, and why
they are important, we first study an example of them for one of the
most simple types of stochastic dynamics, which is Brownian motion
modeled by an ordinary Langevin equation. Along these lines, we then
consider generalized versions of Langevin dynamics exhibiting
anomalous diffusion. For these types of dynamics we check again for
fluctuation relations and in one case obtain a different, new form of
such a formula. We argue that generalized, anomalous fluctuation
relations should be important to understand nonequilibrium
fluctuations in glassy dynamics.

\section{Fluctuation relations}

{\em Fluctuation Relations} (FRs) denote a set of symmetry relations
describing large-deviation properties of the PDFs of statistical
physical observables far from equilibrium. First forms defining one
subset of them, often referred to as {\em Fluctuation Theorems},
emerged from generalizing fluctuation-dissipation relations to
nonlinear stochastic processes \cite{BoKu81a,BoKu81b}. They were then
discovered as generalizations of the Second Law of Thermodynamics for
thermostated dynamical systems, i.e., systems interacting with thermal
reservoirs, in nonequilibrium steady states
\cite{ECM93,EvSe94,GaCo95a,GaCo95b}. Another subset, called {\em
work relations}, generalize a relation between work and free energy,
known from equilibrium thermodynamics, to nonequilibrium situations
\cite{Jar97a,Jar97b}. These two fundamental classes were later on 
amended and generalized by a variety of other FRs from which they can
partially be derived as special cases
\cite{Croo99,HaSa01,Sei05,SaUe10}. Research performed over the past 
ten years has shown that FRs hold for a great variety of systems thus
featuring one of the rare statistical physical principles that is
valid even very far from equilibrium; see, e.g.,
Refs.~\cite{Gall98,EvSe02,Kla06,HaSch07,Seif08,Jarz08,KJJ12} and
further references therein. Many of these relations have meanwhile
been verified in experiments on small systems, i.e., systems on
molecular scales featuring only a limited number of relevant degrees
of freedom \cite{WSM+02,Rit03,BLR05,CJP10,TSUMS10,AlRiRi11}.

So far FRs have mostly been studied for dynamics exhibiting normal
diffusion. This raises the question to which extent the `conventional'
FRs derived for these cases are valid for anomalous dynamics
\cite{ChKl09,KCD12}. Theoretical results for generalized Langevin
equations \cite{BeCo04,OhOh07,MaDh07,CCC08}, L\'evy flights
\cite{ToCo07,ToCo09} and continuous-time random walk models
\cite{EsLi08} as well as computer simulations for glassy dynamics
\cite{Sell09} showed both validity and violations of the various types
of conventional FRs referred to above, depending on the specific type
of anomalous dynamics considered and the nonequilibrium conditions
that have been applied \cite{ChKl09}.

In this section we outline how the two different fields of FRs and
anomalous dynamics can be cross-linked in order to explore to which
extent conventional forms of FRs are valid for anomalous
dynamics. With the term {\em anomalous fluctuation relations} we thus
refer to deviations from conventional forms of FRs, which are due to
anomalous dynamics. Here we focus on generic types of stochastic
anomalous dynamics by only checking {\em transient fluctuation
relations} (TFRs), which describe the approach from a given initial
distribution towards a (non)equilibrium steady state. As a warm-up, we
first derive the conventional TFR for the trivial case of Brownian
motion of a particle moving under a constant external force modeled by
standard Langevin dynamics. We then consider a straightforward
generalization of this type of dynamics in form of long-time
correlated Gaussian stochastic processes. For two fundamental,
different versions of this dynamics we check for the existence of
conventional TFRs under the simple nonequilibrium condition of a
constant external force.

\section{Fluctuation relations for ordinary Langevin dynamics}\label{sec:motiv}

\begin{figure}[t]  
\centerline{\includegraphics[width=5cm]{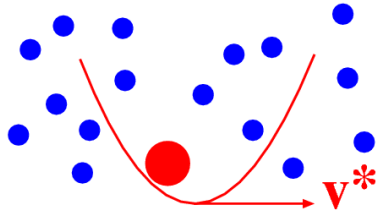}}
\caption{Sketch of a colloidal particle confined within a harmonic
trap that is dragged through water with a constant velocity $v^*$, cf.\ the 
experiment by Wang et al.\ \cite{WSM+02}.}
\label{fig:patrap}
\end{figure}

Consider a particle system evolving from some initial state at time
$t=0$ into a nonequilibrium steady state for $t\to\infty$. A famous
example that has been investigated experimentally \cite{WSM+02} is a
colloidal particle immersed into water and confined by an optical
harmonic trap, see Fig.~\ref{fig:patrap}. The trap is first at rest
but then dragged through water with a constant velocity $v^*$.

The key for obtaining FRs in such systems is to compute the PDF
$\rho(\xi_t)$ of suitably defined dimensionless entropy production
$\xi_t$ over trajectory segments of time length $t$. The goal is to
quantify the asymmetry between positive and negative entropy
production in $\rho(\xi_t)$ for different times $t$ since, as we will
demonstrate in a moment, this relation is intimately related to the
Second Law of Thermodynamics. For a very large class of systems, and
under rather general conditions, it was shown that the following
equation holds \cite{EvSe02,HaSch07,Kurch07}:
\be  
\ln\frac{\rho(\xi_t)}{\rho(-\xi_t)}=\xi_t \label{eq:tfr}\quad .
\ee
Given that here we consider the transient evolution of a system from
an initial into a steady state, this formula became known as the {\em
transient fluctuation relation} (TFR). The left hand side we may call
the fluctuation ratio. Relations exhibiting this functional form have
first been proposed in the seminal work by Evans, Cohen and Morriss
\cite{ECM93}, although in the different situation of considering
nonequilibrium steady states. Such a steady state relation was proved
a few years later on by Gallavotti and Cohen for deterministic
dynamical systems, based on the chaotic hypothesis
\cite{GaCo95a,GaCo95b}. The idea to consider such relations for
transient dynamics was first put forward by Evans and Searles
\cite{EvSe94}.

\begin{figure}[t]
\centerline{\includegraphics[width=6cm,angle=-90]{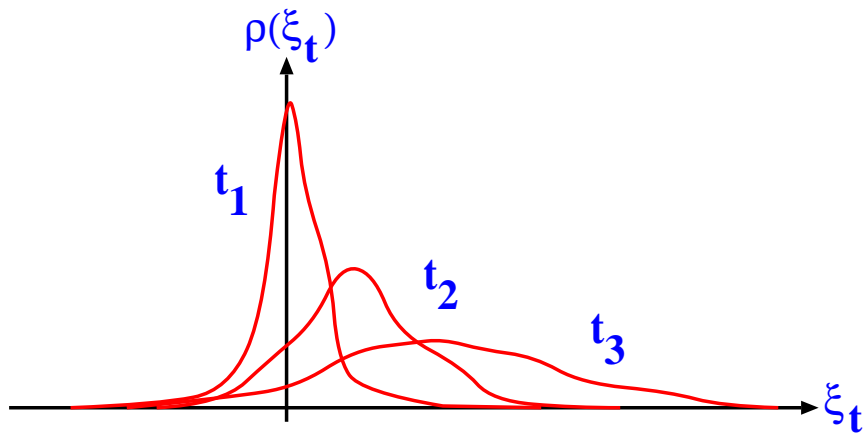}}
\caption{Illustration of the dynamics of the probability density function for
entropy production $\rho(\xi_t)$ for different times $t_1<t_2<t_3$.}
\label{fig:eppdf} 
\end{figure}

Fig.~\ref{fig:eppdf} displays the temporal evolution of the PDF for
entropy production in such a situation. The asymmetry of the evolving
distribution, formalized by the fluctuation relation
Eq.~(\ref{eq:tfr}), is in line with the Second Law of
Thermodynamics. This easily follows from Eq.~(\ref{eq:tfr}) by noting
that
\be
\rho(\xi_t)=\rho(-\xi_t)\exp(\xi_t)\ge\rho(-\xi_t)\;,
\ee
where $\xi_t$ is taken to be positive or zero. Integration from zero
to infinity over both sides of this inequality after multiplication
with $\xi_t$ and defining the ensemble average over the given PDF as
$\langle\ldots\rangle=\int_{-\infty}^{\infty}d\xi_t\;\rho(\xi_t)\ldots$
yields
\be
\langle\xi_t\rangle\ge0\:.\label{eq:frsl}
\ee

As a warm-up, we may first check the TFR for the ordinary overdamped
{\em Langevin equation} \cite{KTH92}
\be
\dot{x}=F+\zeta(t)\quad , \label{eq:ole}
\ee
with a constant external force given by  $F$ and Gaussian white noise
$\zeta(t)$. Note that for sake of simplicity, here we set all the
other constants that are not relevant within this specific context
equal to one. For Langevin dynamics with a constant
force the entropy production $\xi_t$ defined by the heat,
or equivalently the dissipative work, is simply equal to the mechanical work
\cite{vZCo03b}
\be
W_t=Fx(t)\quad .
\ee
It follows that the PDF of entropy production, which here is identical
to the one for the mechanical work, is trivially related to the PDF of
the position $x$ of the Langevin particle via
\be
\rho(W_t)=F^{-1}\varrho(x,t)\quad . \label{eq:whscal}
\ee
This is very convenient, since it implies that all that remains to be
done in order to check the TFR Eq.~(\ref{eq:tfr}) is to solve the
Fokker-Planck equation for the position PDF $\varrho(x,t)$ for a given
initial condition. Here and in the following, we choose $x(0)=0$,
i.e., in terms of position PDFs we start with a delta-distribution at
$x=0$. Note that for ordinary Langevin dynamics in a given
potential, typically the equilibrium density is taken as the initial
density \cite{vZCo03,vZCo03b}. However, since in the following we will
consider dynamics that may not exhibit a simple equilibrium state,
without loss of generality here we make a different choice.

For the ordinary Langevin dynamics Eq.~(\ref{eq:ole}) modeling a
linear Gaussian stochastic process, the position PDF is Gaussian
exhibiting normal diffusion \cite{KTH92,Risk},
\be
\varrho(x,t)=\frac{1}{\sqrt{2\pi\sigma_{x,0}^2}}\exp\left(-\frac{(x-\langle x\rangle )^2}{2\sigma_{x,0}^2}\right)\quad . \label{eq:gauss}
\ee
With the subscript zero we denote ensemble averages 
in case of zero external field. By using the PDF-scaling
Eq.~(\ref{eq:whscal}) and plugging this result into the TFR
Eq.~(\ref{eq:tfr}), we easily derive that the TFR for the work $W_t$
holds if
\be
\langle W_t\rangle =\frac{\sigma_{W_t,0}^2}{2} \quad , \label{eq:fdr1}
\ee
which is nothing else than an example of the {\em
fluctuation-dissipation relation of the first kind} (FDR1)
\cite{Kubo66,KTH92}. We thus arrive at
the seemingly trivial but nevertheless important result that for this
simple Gaussian stochastic process, the validity of FDR1
Eq.~(\ref{eq:fdr1}) implies the validity of the work TFR
Eq.~(\ref{eq:tfr}). For a full analysis of FRs of ordinary Langevin
dynamics we refer to van Zon and Cohen Refs.~\cite{vZCo03,vZCo03b}.

Probably inspired by the experiment of Ref.~\cite{WSM+02}, typically
Langevin dynamics in a harmonic potential moving with a constant
velocity has been studied in the literature
\cite{ZBC05,MaDh07,OhOh07,CCC08}, cf.\ Fig.~\ref{fig:patrap}. Note
that in this slightly more complicated case the (total) work is not
equal to the heat \cite{vZCo03b}. While for the work one recovers the
TFR in its conventional form Eq.~(\ref{eq:tfr}) in analogy to the
calculation above, surprisingly the TFR for heat looks different for
large enough fluctuations. The origin of this phenomenon has been
discussed in detail in Ref.~\cite{vZCo03}, related effects have been
reported in Refs.~\cite{HRS06,ESR05,HaSch07}. However, in the
following we check for another source of deviations from the
conventional TFR Eq.~(\ref{eq:tfr}) that is due to the existence of
microscopic correlations in form of anomalous dynamics.  In order to
illustrate the main ideas it suffices to consider a nonequilibrium
situation simply generated by a constant external force.

\section{Fluctuation relations for anomalous Langevin dynamics}\label{sec:afr}

In our presentation of this section we follow Ref.~\cite{ChKl09},
which may be consulted for further details. Our goal is to check the
TFR Eq.~(\ref{eq:tfr}) for {\em Gaussian stochastic processes}
generating anomalous diffusion. These processes are defined by the
overdamped generalized Langevin equation
\be  
\int_0^t dt'\dot{x}(t')\gamma(t-t')=F+\zeta(t) \label{eq:gle}
\ee
with Gaussian noise $\zeta(t)$ and friction that is modeled with a
memory kernel $\gamma(t)$. By using this equation a stochastic process
can be defined that exhibits normal statistics but with anomalous
memory properties in form of non-Markovian long-time correlated
Gaussian noise.  Equations of this type can be traced back at least to
work by Mori and Kubo around 1965 (see \cite{Kubo66} and further
references therein). They form a class of standard models generating
anomalous diffusion that has been widely investigated, see, e.g.,
Refs.~\cite{KTH92,PWM96,Lutz01}. FRs for this type of dynamics have
more recently been analyzed in
Refs.~\cite{BeCo04,OhOh07,MaDh07,CCC08}. Examples of applications for
this type of stochastic modeling are given by generalized elastic
models \cite{TCK10}, polymer dynamics \cite{Panja10} and biological
cell migration \cite{DKPS08}. We split this class into two specific
cases:\\

{\bf 1. Correlated internal Gaussian noise}\\

We speak of {\em internal} Gaussian noise in the sense that we require
the system to exhibit the {\em fluctuation-dissipation relation of the
second kind} (FDR2) \cite{Kubo66,KTH92}
\be
\langle \zeta(t)\zeta(t')\rangle\sim \gamma(t-t')\:,
\ee
again by neglecting all constants that are not relevant for the main
point we wish to make here. We now consider the specific case that
both the noise and the friction are correlated by a simple power law,
\be
\gamma(t)\sim t^{-\beta}\;,\;0<\beta<1\:.
\ee
Because of the linearity of the generalized Langevin equation
(\ref{eq:gle}) the position PDF must be the Gaussian
Eq.~(\ref{eq:gauss}), and by the scaling of Eq.~(\ref{eq:whscal}) we have
$\rho(W_t)\sim\varrho(x,t)$. It thus remains to solve
Eq.~(\ref{eq:gle}) for mean and variance, which can be done in Laplace
space \cite{ChKl09} yielding {\em subdiffusion},
\be
\sigma_{x,F}^2\sim t^{\beta}\:,
\ee
by preserving the FDR1 Eq.~(\ref{eq:fdr1}). Here and in the following
we denote ensemble averages in case of a non-zero external field with
the subscript $F$. For Gaussian stochastic processes we have seen in
the previous section that the conventional work TFR follows from
FDR1. Hence, for the above power-law correlated internal Gaussian
noise we recover the conventional work TFR Eq.~(\ref{eq:tfr}).\\

{\bf 2. Correlated external Gaussian noise}\label{sec:cegn}\\

As a second case, we consider the overdamped generalized Langevin
equation
\be
\dot{x}=F+\zeta(t)\label{eq:gle2}\:,
\ee
which represents a special case of Eq.~(\ref{eq:gle}) with a
memory kernel modeled by a delta-function. Again we use correlated
Gaussian noise defined by the power law
\be
\langle\zeta(t)\zeta(t')\rangle\sim |t-t'|^{-\beta}\;,\;0<\beta<1\:,
\ee
which one may call {\em external}, because in this case we do not
postulate the existence of FDR2. The position PDF is again Gaussian,
and as before $\rho(W_t)\sim\varrho(x,t)$. However, by solving the
Langevin equation along the same lines as in the previous case, here
one obtains {\em superdiffusion} by breaking FDR1,
\begin{figure}[t]  
\centerline{\includegraphics[width=10cm]{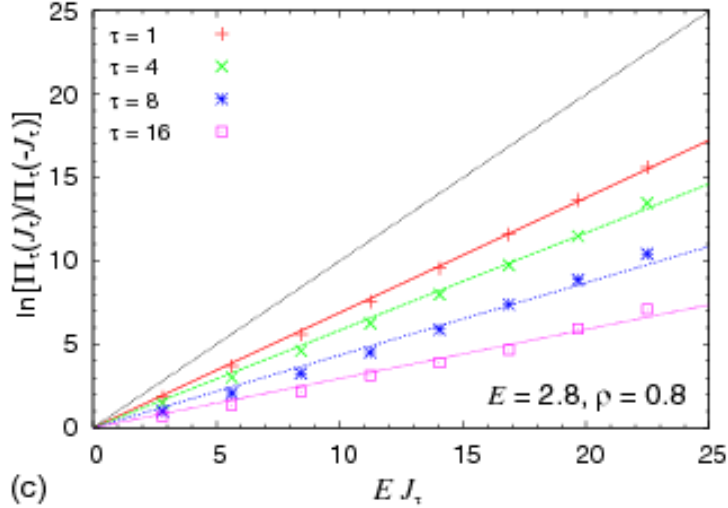}}
\caption{The fluctuation ratio 
$\ln(\Pi_{\tau}(J_{\tau})/\Pi_{\tau}(J_{\tau}))$ for the entropy
production $W_{\tau}=EJ_{\tau}$ with particle current $J_{\tau}$ and
field strength $E$ for particle density $\rho$ at different times
$\tau$. The full line, with slope one, displays the result of the
conventional FR Eq.~(\ref{eq:tfr}) in a nonequilibrium steady state.
The figure is from Ref.~\cite{Sell09}.}
\label{fig:sellafr}
\end{figure}
\be
\langle W_t\rangle\sim t\quad , \quad \sigma_{W_t,F}^2\sim t^{2-\beta}\:.
\ee
Calculating the fluctuation ratio, i.e., the left hand side of
Eq.~(\ref{eq:tfr}), from these results yields
the {\em anomalous work TFR}
\be
\ln\frac{\rho(W_t)}{\rho(-W_t)}=C_{\beta} t^{\beta-1}W_t \quad 0<\beta<1\:,\label{eq:afr}
\ee
where $C_{\beta}$ is a constant that depends on physical parameters
\cite{ChKl09}. Comparing this equation with the conventional form of
the TFR Eq.~(\ref{eq:tfr}) one observes that the fluctuation ratio is
still linear in $W_t$ thus exhibiting an exponential large deviation
form \cite{ToCo09}. However, there are two important deviations: (1)
the slope of the fluctuation ratio as a function of $W_t$ is not equal
to one anymore, and in particular (2) it decreases with time. We may
thus classify Eq.~(\ref{eq:afr}) as a {\em weak violation of the
conventional TFR}.

We remark that for driven glassy systems FRs have already been
obtained displaying slopes that are not equal to one. Within this
context it has been suggested to capture these deviations from one by
introducing the concept of an `effective temperature'
\cite{Sell98,ZRA05,ZBC05}. As far as the time dependence of the
coefficient is concerned, such behavior has recently been observed in
computer simulations of a paradigmatic two dimensional lattice gas
model generating glassy dynamics \cite{Sell09}. Fig.~\ref{fig:sellafr}
shows the fluctuation ratio as a function of the entropy production at
different times $\tau$ as extracted from computer simulations of this
model, where the PDF has first been relaxed into a nonequilibrium
steady state. It is clearly seen that the slope decreases with time,
which is in line with the prediction of the anomalous TFR
Eq.~(\ref{eq:afr}). To which extent the nonequilibrium dynamics of
this lattice gas model can be mapped onto the generalized Langevin
equation Eq.~(\ref{eq:gle2}) is an open question.

In summary, for Gaussian stochastic processes with correlated noise
the existence of FDR2 implies the existence of FDR1, and FDR1 in turn
implies the existence of work TFR in conventional form. That is, the
conventional work TFR holds for internal noise. However, there is a
weak violation of the conventional form in case of external noise
yielding a pre-factor that is not equal to one and in particular
depends on time.

\chapter{Anomalous dynamics of biological cell migration}\label{sec:cell}

In order to illustrate the importance of anomalous dynamics for
realistic situations, in this final section of our book chapter we
discuss experiments and theory about the migration of single
biological cells crawling on surfaces as an example. Here we focus on
cells in an equilibrium situation, i.e., not moving under the
influence of any external gradients or fields. This case is
investigated by extracting results for the MSD and for the position
PDFs from experimental data. We then show how the experimental results
can be understood by a mathematical model in form of a fractional
Klein-Kramers equation. As far as MSD and velocity autocorrelation
function are concerned, this equation bears some similarity to a
generalized Langevin equation that is of the same type as the one that
has been discussed in Section~\ref{sec:cegn}. Our presentation in this
section is based on Ref.~\cite{DKPS08}.

\section{Cell migration}

\begin{figure}[t]  
\centerline{\includegraphics[height=10cm,angle=-90]{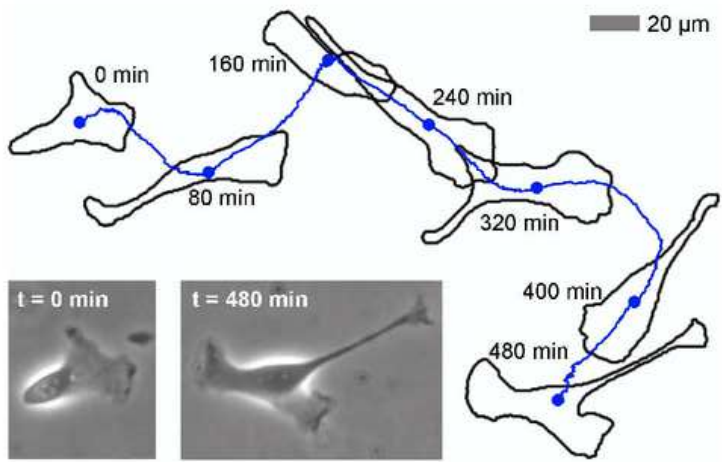}}
\caption{Overlay of a biological cell migrating {\em in vitro}
on a substrate. The cell frequently changes its shape and direction
during migration, as is shown by several cell contours extracted
during the migration process. The inset displays phase contrast images
of the cell at the beginning and to the end of its migration process
\cite{DKPS08}.}
\label{fig:cell}
\end{figure}

Nearly all cells in the human body are mobile at a given time during
their life cycle. Embryogenesis, wound-healing, immune defense and the
formation of tumor metastases are well known phenomena that rely on
cell migration \cite{LaHo96,LaSi09,FrWo10}. Fig.~\ref{fig:cell}
depicts the path of a single biological cell crawling on a substrate
measured in an {\em in vitro} experiment \cite{DKPS08}. At first
sight, the path looks like the trajectory of a Brownian particle
generated, e.g., by the ordinary Langevin dynamics of
Eq.~(\ref{eq:ole}). On the other hand, according to Einstein's theory
of Brownian motion a Brownian particle is {\em passively} driven by
collisions from the surrounding fluid molecules, whereas biological
cells move {\em actively} by themselves converting chemical into
kinetic energy. This raises the question whether the random-looking
paths of crawling biological cells can really be understood in terms
of simple Brownian motion
\cite{DuBr87,SLW91} or whether more advanced concepts of dynamical
modeling have to be applied \cite{HLCC94,URGS01,LNC08,TSYU08,BBFB10}.

\section{Experimental results and statistical analysis}

The cell migration experiments that we now discuss have been performed
on two types of tumor-like migrating {\em transformed renal epithelial
Madin Darby canine kidney (MDCK-F)} cell strains: wild-type ($NHE^+$)
and NHE-deficient ($NHE^-$) cells. Here $NHE^+$ stands for a molecular
sodium hydrogen exchanger that either is present or deficient. It can
thus be checked whether this microscopic exchanger has an influence on
cell migration, which is a typical question asked by cell
physiologists. The cell diameter is about 20-50$\mu$m and the mean
velocity of the cells about $1\mu$m/min. Cells are driven by active
protrusions of growing actin filaments ({\em lamellipodial dynamics})
and coordinated interactions with myosin motors and dynamically
re-organizing cell-substrate contacts. The leading edge dynamics of a
polarized cell proceeds at the order of seconds. Thirteen cells were
observed for up to 1000 minutes. Sequences of microscopic phase
contrast images were taken and segmented to obtain the cell boundaries
shown in Fig.~\ref{fig:cell}; see Ref.~\cite{DKPS08} for full details
of the experiments.

\begin{figure}[t]  
\centerline{\includegraphics[height=14cm,angle=-90]{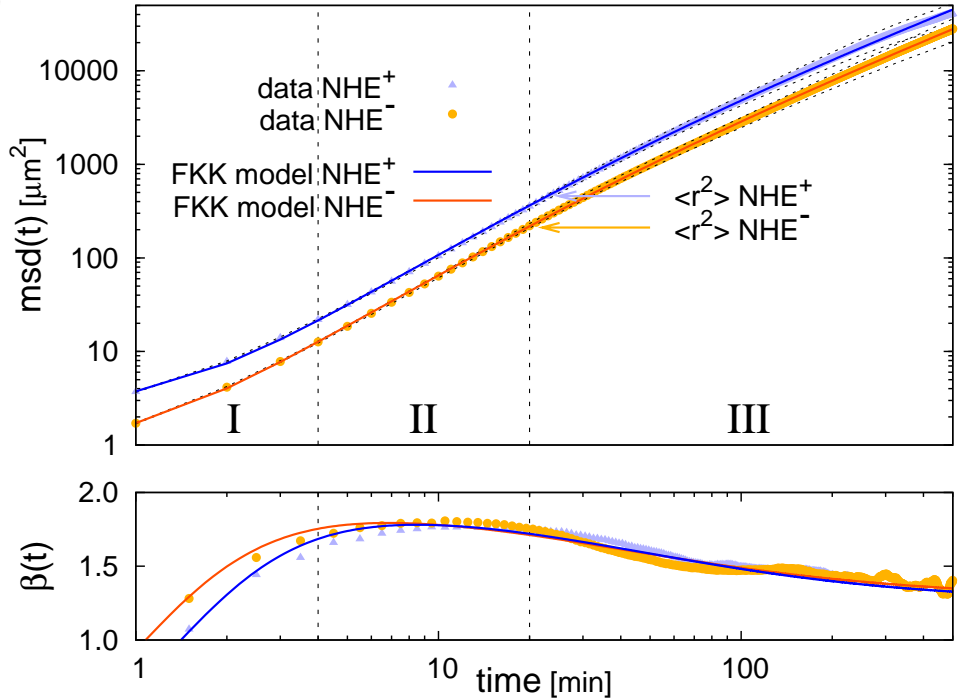}}
\caption{Upper part: Double-logarithmic plot of the mean square 
displacement (MSD) as a function of time. Experimental data points for
both cell types are shown by symbols. Different time scales are marked
as phases I, II and III as discussed in the text. The solid lines
represent fits to the MSD from the solution of our model, see
Eq.~(\ref{eq:msdn}). All parameter values of the model are given in
\cite{DKPS08}. The dashed lines indicate the uncertainties of the MSD
values according to Bayes data analysis. Lower part: Logarithmic
derivative $\beta(t)$ of the MSD for both cell types as defined by
Eq.~(\ref{eq:lder}).}
\label{fig:cell_msd}
\end{figure}

According to the Langevin description of Brownian motion outlined in
Section~\ref{sec:motiv}, Brownian motion is characterized by a MSD
$\sigma_{x,0}^2(t)\sim t\:(t\to\infty)$ designating normal
diffusion. Fig.~\ref{fig:cell_msd} shows that both types of cells
behave differently: First of all, MDCK-F $NHE^-$ cells move less
efficiently than $NHE^+$ cells resulting in a reduced MSD for all
times. As is displayed in the upper part of this figure, the MSD of
both cell types exhibits a crossover between three different dynamical
regimes. These three phases can be best identified by extracting the
time-dependent exponent $\beta$ of the MSD $\sigma_{x,0}^2(t)\sim
t^{\beta}$ from the data, which can be done by using the logarithmic
derivative \be \beta(t)=\frac{d\ln msd(t)}{d \ln t} \quad
. \label{eq:lder} \ee The results are shown in the lower part of
Fig.~\ref{fig:cell_msd}. Phase I is characterized by an exponent
$\beta(t)$ roughly below $1.8$. In the subsequent intermediate phase
II, the MSD reaches its strongest increase with a maximum exponent
$\beta$. When the cell has approximately moved beyond a square
distance larger than its own mean square radius (indicated by arrows
in the figure), $\beta(t)$ gradually decreases to about $1.4$. Both
cell types therefore do not exhibit normal diffusion, which would be
characterized by $\beta(t)\to 1$ in the long time limit, but move
anomalously, where the exponent $\beta>1$ indicates superdiffusion.

\begin{figure}[t]
\centerline{\hspace*{-0.5cm}\includegraphics[height=11cm]{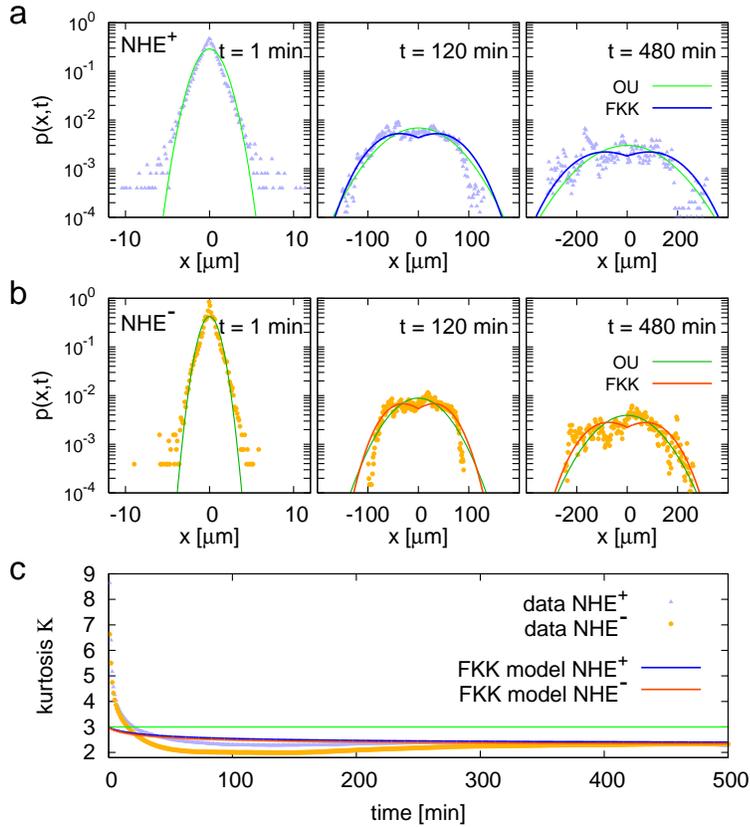}}
\caption{Spatio-temporal probability distributions $P(x,t)$. (a),(b):
Experimental data for both cell types at different times in
semilogarithmic representation. The dark lines, labeled FKK, show the
long-time asymptotic solutions of our model Eq.~(\ref{eq:fkk}) with
the same parameter set used for the MSD fit. The light lines, labeled
OU, depict fits by the Gaussian distributions Eq.~(\ref{eq:gauss})
representing Brownian motion. For $t=1$ min both $P(x,t)$ show a
peaked structure clearly deviating from a Gaussian form. (c) The
kurtosis $\kappa(t)$ of $P(x,t)$, cf.\ Eq.~(\ref{eq:kurt}), plotted as
a function of time saturates at a value different from the one of
Brownian motion (line at $\kappa=3$). The other two lines represent
$\kappa(t)$ obtained from the model Eq.~(\ref{eq:fkk}) \cite{DKPS08}.}
\label{fig:cell_pdf} \end{figure}

We next study the PDF of cell positions. Since no correlations between
$x$ and $y$ positions could be found, it suffices to restrict
ourselves to one dimension. Fig.~\ref{fig:cell_pdf} (a), (b) reveals
the existence of non-Gaussian distributions at different times. The
transition from a peaked distribution at short times to rather broad
distributions at long times suggests again the existence of distinct
dynamical processes acting on different time scales. The shape of
these distributions can be quantified by calculating the {\em
kurtosis}
\be
\kappa(t):=\frac{\langle x^4(t)\rangle}{\langle x^2(t)\rangle^2} \quad ,\label{eq:kurt}
\ee
which is displayed as a function of time in Fig.~\ref{fig:cell_pdf}
(c). For both cell types $\kappa(t)$ rapidly decays to a constant that
is clearly below three in the long time limit. A value of three would
be the result for the spreading Gaussian distributions characterizing
Brownian motion. These findings are another strong manifestation of
the anomalous nature of cell migration.

\section{Stochastic modeling}

We now present the stochastic model that we have used to reproduce the
experimental data yielding the fit functions shown in the previous two
figures. The model is defined by the {\em fractional Klein-Kramers
equation} \cite{BaSi00} \be
\pard{\varrho}{t}=-\pard{}{x}\left[v\varrho\right]+
\pard{^{1-\alpha}}{t^{1-\alpha}}\gamma_{\alpha}\left[\pard{}{v}v+
v_{th}^2\pard{^2}{v^2}\right]\varrho \quad
\:,\:0<\alpha<1\:. \label{eq:fkk}
\ee 
Here $\varrho=\varrho(x,v,t)$ is the PDF depending on time $t$,
position $x$ and velocity $v$ in one dimension, $\gamma_{\alpha}$ is a
friction term and $v_{th}^2=k_B T/M$ stands for the thermal velocity
squared of a particle of mass $M=1$ at temperature $T$, where $k_B$ is
Boltzmann's constant. The last term in this equation models diffusion
in velocity space. In contrast to Fokker-Planck equations, this
equation features time evolution both in position and velocity
space. What distinguishes this equation from an ordinary Klein-Kramers
equation, the most general model of Brownian motion \cite{Risk}, is
the presence of the Riemann-Liouville fractional derivative of order
$1-\alpha$
\be
\frac{\partial^{1-\alpha}}{\partial t^{1-\alpha}}  \varrho = 
\frac{\partial}{\partial t}\left[\frac{1}{\Gamma (\alpha)} \int_0^t dt^{\prime }\frac{\varrho(t^{\prime})}{(t-t^{\prime })^{1-\alpha}}\right] \label{eq:rlfd}
\ee
in front of the terms in square brackets. Note that for $\alpha=1$ the
ordinary Klein-Kramers equation is recovered. The analytical solution
of this equation for the MSD has been calculated in Ref.~\cite{BaSi00}
to
\be
\sigma_{x,0}^2(t)=2v_{th}^2t^2E_{\alpha,3}(-\gamma_{\alpha} t^{\alpha})
\quad \to\quad 2\frac{D_{\alpha}t^{2-\alpha}}{\Gamma(3-\alpha)}\quad
(t\to\infty) \label{eq:fkkmsd} 
\ee 
with $D_{\alpha}=v_{th}^2/\gamma_{\alpha}$ and the {\em
two-parametric} or {\em generalized Mittag-Leffler function} (see,
e.g., Chapter 4 of Ref.~\cite{KRS08} and Refs.~\cite{GoMa97,Podl99})
\be
E_{\alpha,\beta}(z)=\sum_{k=0}^{\infty}\frac{z^k}{\Gamma(\alpha
k+\beta)}\:,\:\alpha\:,\:\beta>0\:,\:z\in\mathbb{C} \quad .
\ee 
Note that $E_{1,1}(z)=\exp(z)$, hence $E_{\alpha,\beta}(z)$ is a
generalized exponential function. We see that for long times
Eq.~(\ref{eq:fkkmsd}) yields a power law, which reduces to the
long-time Brownian motion result in case of $\alpha=1$. 

In view of the experimental data shown in Fig.~\ref{fig:cell_msd},
Eq.~(\ref{eq:fkkmsd}) was amended by including the impact of random
perturbations acting on very short time scales for which we take
Gaussian white noise of variance $\eta^2$. This leads to \cite{MFK02}
\be
\sigma_{x,0;noise}^2(t)=\sigma_{x,0}^2(t)+2\eta^2 \quad
. \label{eq:msdn} 
\ee 
The second term mimicks both measurement errors and fluctuations of
the cell cytoskeleton. In case of the experiments with MDCK-F
cells \cite{DKPS08}, the value of $\eta$ can be extracted from the
experimental data and is larger than the estimated measurement
error. Hence, this noise must largely be of a biological nature and
may be understood as being generated by microscopic fluctuations of
the lamellipodia in the experiment.

The analytical solution of Eq.~(\ref{eq:fkk}) for $\varrho(x,v,t)$ is
not known, however, for large friction $\gamma_{\alpha}$ this equation
boils down to a fractional diffusion equation for which $\varrho(x,t)$
can be calculated in terms of a Fox function \cite{SchnWy89}. The
experimental data in Figs.~\ref{fig:cell_msd} and \ref{fig:cell_pdf}
was then fitted consistently by using the above solutions with the
four parameters $v_{th}^2, \alpha,\gamma$ and $\eta^2$ in Bayesian
data analysis \cite{DKPS08}.

In summary, by statistical analysis of experimental data we have shown
that the equilibrium migration of the biological cells under
consideration is anomalous. Related anomalies have also been observed
for other types of migrating cells
\cite{HLCC94,URGS01,LNC08,TSYU08,BBFB10}. Our experimental results
are coherently reproduced by a mathematical model in form of a
stochastic fractional equation. We now elaborate on possible physical
and biological interpretations of our findings.

First of all, we remark that the solutions of Eq.~(\ref{eq:fkk}) for
both the MSD and the velocity autocorrelation function match precisely
to the solutions of the generalized Langevin equation \cite{Lutz01}
\be 
\dot{v}=-\int_0^tdt'\:\gamma(t-t')
v(t')+\:\xi(t) \label{eq:ugle} \:.
\ee 
Here $\xi(t)$ holds for Gaussian white noise and $\gamma(t)\sim
t^{-\alpha}$ for a time-dependent friction coefficient with a power
law memory kernel, which alternatively could be written by using a
fractional derivative \cite{Lutz01}. For $\gamma(t)\sim\delta(t)$ the
ordinary Langevin equation is recovered. Note that the position PDF
generated by this equation is Gaussian in the long time limit and thus
does not match to the one of the fractional Klein-Kramers equation
Eq.~(\ref{eq:fkk}). However, alternatively one could sample from a
non-Gaussian $\xi(t)$ to generate a non-Gaussian position
PDF. Strictly speaking, despite equivalent MSD and velocity
correlations Eqs.~(\ref{eq:fkk}) and (\ref{eq:ugle}) define different
classes of anomalous stochastic processes. The precise cross-links
between the Langevin description and the fractional Klein-Kramers
equation are subtle \cite{EFJK07} and to some extent still
unknown. The advantage of Eq.~(\ref{eq:ugle}) is that it allows more
straightforwardly a possible biophysical interpretation of the origin
of the observed anomalous MSD and velocity correlations, at least
partially, in terms of the existence of a memory-dependent friction
coefficient. The latter, in turn, might be explained by anomalous
rheological properties of the cell cytoskeleton, which consists of a
complex biopolymer gel
\cite{SSGMBK07}.

Secondly, what could be the possible biological significance of the
observed anomalous cell migration? There is an ongoing debate about
whether biological organisms such as, e.g., albatrosses, marine
predators and fruit flies have managed to mimimize the search time for
food in a way that matches to optimizing search strategies in terms of
stochastic processes; see Refs.~\cite{BLMV11,VLRS11} and further
references therein. In particular, it has been argued that L\'evy
flights are superior to Brownian motion in order to find sparsely,
randomly distributed, replenishing food sources
\cite{BLMV11}. However, it was also shown that in other situations
{\em intermittent dynamics} is more efficient than pure L\'evy motion
\cite{BLMV11}.  For our cell experiment, both the experimental data
and the theoretical modeling suggest that there exists a slow
diffusion on short time scales, whereas the long-time motion is much
faster, which resembles intermittency as discussed in
Ref.~\cite{BLMV11}.  Hence, the results on anomalous cell migration
presented above might be biologically relevant in view of suitably
optimized foraging strategies.

\chapter{Summary}\label{sec:summ}

This chapter highlighted some fundamental aspects of anomalous
dynamics: The scene was set by section~\ref{sec:cad}, which reviewed
basic ideas of weak chaos by establishing crosslinks to infinite
ergodic theory. This branch of ergodic theory provides a rigorous
mathematical approach to study weakly chaotic dynamical systems. In
particular, we proposed suitable definitions of generalized chaos
quantities assessing weakly chaotic dynamics by yielding a generalized
version of Pesin's theorem. We also outlined a generalized hierarchy
of chaos on the basis of different functional forms of the dispersion
exhibited by nearby trajectories of a deterministic dynamical
system. In section~\ref{sec:anodif} we related these concepts to the
problem of anomalous diffusion by spatially extending our previously
discussed simple map model. Applying stochastic continuous time random
walk theory to this model in comparison to computer simulations, we
learned about an intricate dynamical phase transition between normal
and anomalous diffusion, governed by multiplicative logarithmic
corrections in the mean square displacement. We also derived a
fractional diffusion equation that reproduced the subdiffusive
diffusive dynamics of this model on coarse scales. The subsequent
section~\ref{sec:afrc} elaborated on fluctuation relations, here
understood as a large-deviation symmetry property of the work
probability distributions generated by a given stochastic dynamics far
from equilibrium. We familiarized ourselves with the conventional form
of transient work fluctuation relations derived from standard Langevin
dynamics before exploring anomalous generalizations of Langevin
equations. One of them reproduced the conventional form of fluctuation
relations, whereas the other one yielded a generalized, anomalous
fluctuation relation. The precise form of the resulting fluctuation
relation appeared to be intimately releated to whether or not
fluctuation-dissipation relations are broken. In our final main
section we related our previous theoretical ideas to the experimental
problem of studying biological cell migration. By extracting the mean
square displacement and the position probability distributions from
experimental data, we found that the dynamics exhibited by these cells
was anomalous, showing different behavior on different time scales, by
eventually yielding superdiffusion for long times. On the basis of
these experimental results we suggested a stochastic theoretical model
of cell migration in form of a fractional Klein-Kramers equation,
which coherently reproduced our experimental findings.

In summary, we traversed quite an anomalous scientific landscape of
different but related topics: Starting from simple deterministic maps
and their ergodic theory description we switched to basics of
anomalous stochastic processes, studied both normal and anomalous
stochastic fluctuations very far from equilibrium in terms of Langevin
dynamics by ending up with anomalously crawling biological cells. We
thus meant to illustrate the third column displayed in the very first
Fig.~\ref{fig:bigp} of the introduction, by also explaining the title
of this contribution. Within a larger scientific context, one may
consider our disucssion as an indication that a novel theory of
anomalous nonequilibrium processes is presently emerging. In contrast
to standard nonequilibrium statistical mechanics, this dynamics is
inherently non-stationary, due to the weak chaos by which it is
generated. This mechanism leads to important physical consequences
like anomalous transport, which can be tested in experiments. On the
side of theoretical physics this approach asks for further
generalizations of recently developed fundamental concepts, perhaps
leading to a weakly chaotic hypothesis, the identification of the
physically relevant measures characterizing such systems, and to
deriving experimentally measurable consequences such as
generalizations of ordinary large deviation properties and fluctuation
relations. However, these questions also motivate further mathematical
work in upcoming directions of infinite ergodic theory to provide a
formal framework and rigorous results for parts of the physical
theory.\\

{\bf Acknowledgements:}\\
Each of the four sections reflects the collaboration with colleagues,
without whom the work presented here would not have been possible. The
second section benefitted very much from discussion with
R.Zweim{\"u}ler, whom the author thanks very much for a lot of
mathematical insight into aspects of infinite ergodic
theory. Particularly, the author is indebted to his former postdoc
P.Howard, who did brilliant work on calculating generalized chaos
quantities for the Pomeau-Manneville map. Regarding the third section,
credit goes to his former PhD student N.Korabel for joint work that
formed part of his PhD thesis. A.V.Chechkin significantly contributed
to the same section, as well as performing major research on the topic
coverd by the fourth one. The author is deeply indebted to him for his
long-term collaboration on anomalous stochastic processes. P.Dieterich
was the driving force behind the project reviewed in the fifth
section. The author thanks him for much insight into the biophysical
aspects of biological cell migration. Finally, he wishes to thank the
editors of this book for their patience with this book chapter.

\newpage


\end{document}